\newcommand*{\ATLASLATEXPATH}{}
\documentclass[PAPER, atlasdraft=false, cernpreprint, texlive=2016, UKenglish, LANGEDIT=true, LANGSHOW=false]{\ATLASLATEXPATH atlasdoc}
\LEcontact{Steve Lloyd s.l.lloyd@qmul.ac.uk}
 
\usepackage[subcaption]{\ATLASLATEXPATH atlaspackage}
\usepackage[block=space]{\ATLASLATEXPATH atlasbiblatex}
 
\usepackage{\ATLASLATEXPATH atlasphysics}
\usepackage{\ATLASLATEXPATH atlasbsm}

\graphicspath{{logos/}{figures/}}
 
\usepackage{ANA-SUSY-2018-32-PAPER-defs}
\usepackage{multirow}
\usepackage{graphicx}
\usepackage{epstopdf}
\usepackage{rotating}
\usepackage{footnote}
\makesavenoteenv{tabular}
\makesavenoteenv{table}
\newenvironment{DIFnomarkup}{}{}

\AtlasTitle{Search for electroweak production of charginos and sleptons decaying into final states with two leptons and missing transverse momentum in $\sqrt{s}=13$~\TeV\ $pp$ collisions using the ATLAS detector}
 
\PreprintIdNumber{CERN-EP-2019-106}
\AtlasAbstract{
A search for the electroweak production of charginos and sleptons decaying into final states with two electrons or muons is presented. The analysis is based on 139~fb$^{-1}$ of proton--proton collisions recorded by the ATLAS detector at the Large Hadron Collider at $\sqrt{s}=13$~\TeV. Three $R$-parity-conserving scenarios where the lightest neutralino is the lightest supersymmetric particle are considered: the production of chargino pairs with decays via either $W$  bosons or sleptons, and the direct production of slepton pairs. The analysis is optimised for the first of these scenarios, but the results are also interpreted in the others.
No significant deviations from the Standard Model expectations are observed and limits at 95\% confidence level are set on the masses of relevant supersymmetric particles in each of the scenarios. For a massless lightest neutralino, masses up to 420~\GeV\ are excluded for the production of the lightest-chargino pairs assuming $W$-boson-mediated decays and up to 1~\TeV\ for slepton-mediated decays, whereas for slepton-pair production masses up to 700 \GeV\ are excluded assuming three generations of mass-degenerate sleptons.
}
 
\author{The ATLAS Collaboration}
\AtlasJournalRef{Eur. Phys. J. C 80 (2020) 123}
\AtlasDOI{10.1140/epjc/s10052-019-7594-6}
 
\AtlasRefCode{SUSY-2018-32}

\AtlasCoverSupportingNote{Search for direct chargino pair production with $W$-boson mediated decays in events with two leptons and missing transverse momentum in the final state at $\boldmath{\sqrt{s}=13\,}$TeV with the ATLAS detector}{https://cds.cern.ch/record/2637338}
 
\AtlasCoverCommentsDeadline{22 May 2019}
 
\AtlasCoverAnalysisTeam{S. Carr\`a, E. Gorini, E. Gramstad, F.G. Gravili, T. Lari, L. Longo, A. Mirto, H. Oppen, H. Pacey, M. Primavera, F. Ould-Saada, M. Reale, H. Sandaker, A. Ventura, S. Williams}

\AtlasCoverEdBoardMember{EdBoard~Chair~(A. De Santo)}
\AtlasCoverEdBoardMember{EdBoard~Member~(V. Mitsou)}
\AtlasCoverEdBoardMember{EdBoard~Member~(O. Rifki)}
 
\AtlasCoverEgroupEditors{atlas-SUSY-2018-32-editors@cern.ch}
 
\AtlasCoverEgroupEdBoard{atlas-SUSY-2018-32-editorial-board@cern.ch}
 
\hypersetup{pdftitle={ATLAS document},pdfauthor={The ATLAS Collaboration}}

\begin{document}
 
\maketitle
 
\tableofcontents
 
\section{Introduction}
\label{sec:intro}
 
Weak-scale supersymmetry (SUSY)~\cite{Golfand:1971iw,Golfand:1971iw_1,Volkov:1973ix,Wess:1974tw,Wess:1974jb,Ferrara:1974pu,Salam:1974ig} is a theoretical extension to the Standard Model (SM) that, if realised in nature, would solve the hierarchy problem~\cite{Sakai:1981gr,Dimopoulos:1981yj,Ibanez:1981yh,Dimopoulos:1981zb} through the introduction of a new fermion (boson) supersymmetric partner for each boson (fermion) in the SM\@. In SUSY models that conserve $R$-parity~\cite{Farrar:1978xj}, SUSY particles (sparticles) must be produced in pairs. The lightest supersymmetric particle (LSP) is stable and weakly interacting, thus potentially providing a viable candidate for dark matter~\cite{Goldberg:1983nd,Ellis:1983ew}. Due to its stability, any LSP produced at the Large Hadron Collider (LHC) would escape detection and give rise to momentum imbalance in the form of missing transverse momentum ($\mathbf p^\mathrm{miss}_\mathrm{T}$) in the final state, which can be used to discriminate SUSY signals from the SM background.
 
The superpartners of the SM Higgs boson and the electroweak gauge bosons, known as the higgsinos, winos and binos, are collectively labelled as electroweakinos. They mix to form chargino ($\tilde{\chi}_i^\pm, i=1,2$) and neutralino ($\tilde{\chi}_j^{0}, j=1,2,3,4$) mass eigenstates where the labels $i$ and $j$ refer to states of increasing mass.
 
Sparticle production cross-sections at the LHC are highly dependent on the sparticle masses as well as on the production mechanism. The coloured sparticles (squarks and gluinos) are strongly produced and have significantly larger production cross-sections than non-coloured sparticles of equal masses, such as the sleptons (superpartners of the SM leptons) and the electroweakinos.
If gluinos and squarks were much heavier than low-mass electroweakinos, 
then SUSY production at the LHC would be dominated by direct electroweakino production. The latest ATLAS and CMS limits on squark and gluino production~\cite{squarks:2018,SUSY-2016-10,SUSY-2016-13,SUSY-2016-14,CMS-SUS-16-033,CMS-SUS-16-036,CMS-SUS-16-050,CMS-SUS-16-038,CMS-SUS-17-006} extend well beyond the \TeV\ scale, thus making electroweak production of sparticles a promising and important probe in searches for SUSY at the LHC.
 
This paper presents a search for the electroweak production of charginos and sleptons decaying into final states with two charged leptons (electrons and/or muons) using
139~fb$^{-1}$ of proton--proton collision data recorded by the ATLAS detector at the LHC at $\sqrt{s}=13$~\TeV.
The analysis is optimised to target the direct production of $\tilde{\chi}_1^+\tilde{\chi}_1^-$, where each chargino decays into the LSP $\tilde{\chi}_1^{0}$ and an on-shell $W$~boson.  
Signal events are characterised by the presence of exactly two isolated leptons ($e$, $\mu$)
with opposite electric charge, and significant $\mathbf p^\mathrm{miss}_\mathrm{T}$ (the magnitude of which is referred to as $\MET$), expected from neutrinos and LSPs in the final states. The same analysis strategy is also applied to two other  searches.
One of them is the search for the direct production of $\tilde{\chi}_1^{+} \tilde{\chi}_1^{-}$, where each chargino decays into a slepton (charged slepton $\tilde{\ell}$ or sneutrino $\tilde{\nu}$) via the emission of a lepton (neutrino $\nu$ or charged lepton $\ell$) and the slepton itself decays into a lepton and the LSP\@. The other one is the search for the direct pair production of sleptons where each slepton decays into a lepton and the LSP\@.
 
The search described here significantly extends the areas of the parameter space beyond those excluded by previous 
searches by ATLAS~\cite{SUSY-2013-11,SUSY-2016-24} and CMS~\cite{CMS-SUS-13-006,CMS-SUS-14-002,CMS-SUS-16-039,CMS-SUS-17-004,CMS-SUS-17-009,CMS-SUS-17-010,} in the same channels.
 
After a description of the considered SUSY scenarios in Section~\ref{sec:scenarios} and of the ATLAS detector in Section~\ref{sec:detector}, the data and simulated Monte Carlo (MC) samples used in the analysis are detailed in Section~\ref{sec:mcsamples}. Section~\ref{sec:objevent} and Section~\ref{sec:SRs} present the event reconstruction and the search strategy. The SM background estimation and the systematic uncertainties are discussed in Section~\ref{sec:backgrounds} and Section~\ref{sec:systematics}, respectively. Finally, the results and their interpretations are reported in Section~\ref{sec:results}. Section~\ref{sec:conclusions} summarises the conclusions.

\section{SUSY scenarios}
\label{sec:scenarios}
\begin{figure}[htb!]
\centering
\begin{subfigure}{0.30\linewidth}
\includegraphics[width=\linewidth]{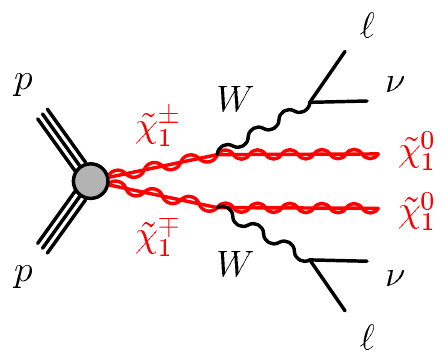}
\caption{}
\end{subfigure}
\begin{subfigure}{0.30\linewidth}
\vspace{-1.6cm}
\includegraphics[width=\linewidth]{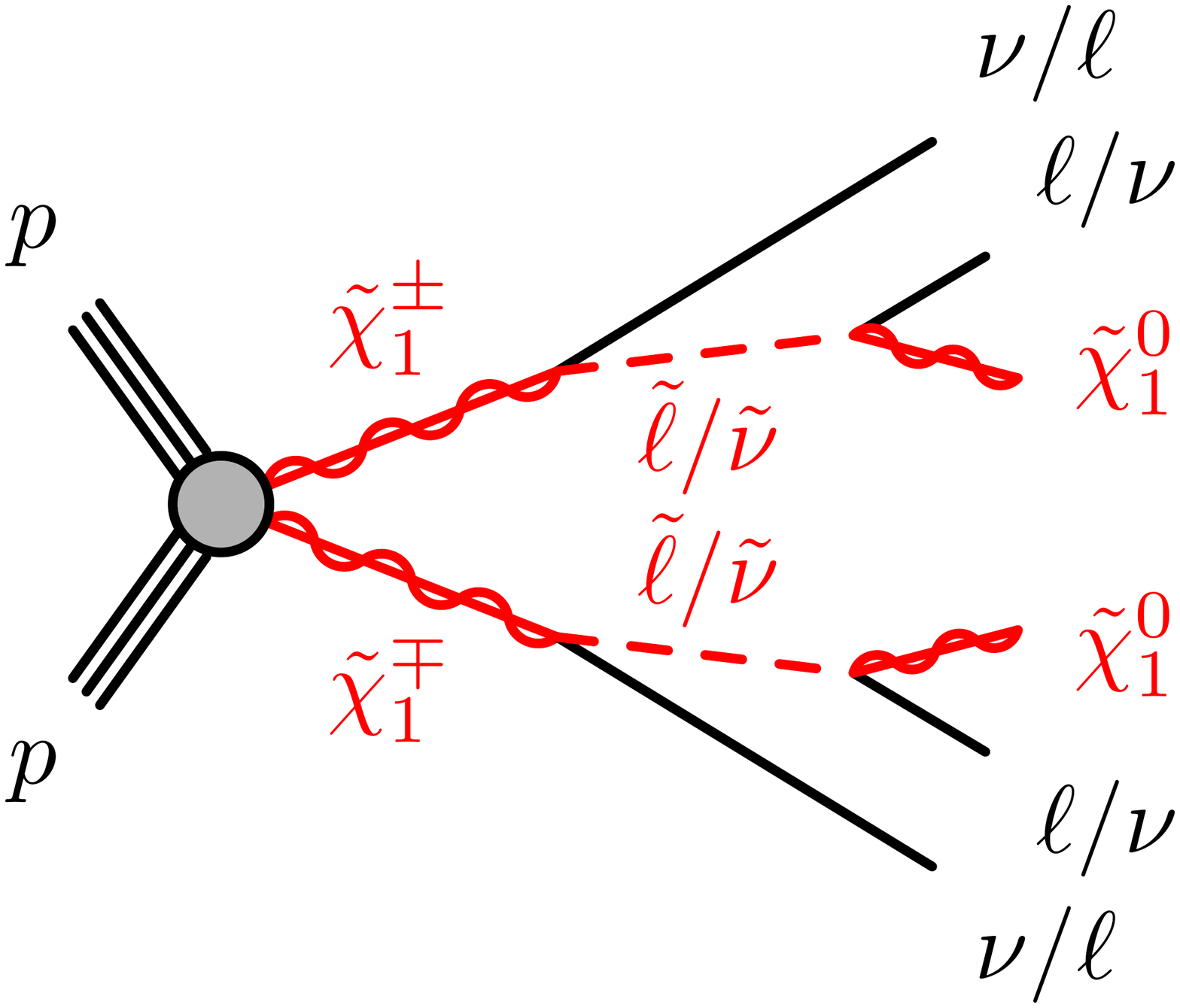}
\vspace{-1.6cm}
\caption{}
\end{subfigure}
\begin{subfigure}{0.30\linewidth}
\includegraphics[width=\linewidth]{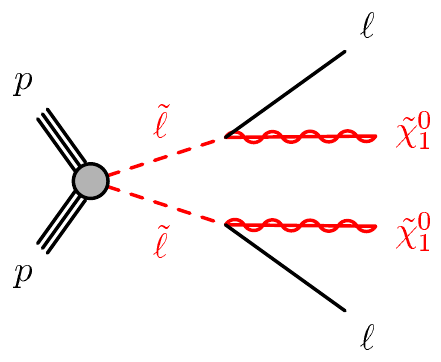}
\caption{}
\end{subfigure}
\caption{Diagrams of the supersymmetric models considered, with two leptons and weakly interacting particles in the final state: (a) $\tilde{\chi}_1^+\tilde{\chi}_1^-$ production with $W$-boson-mediated decays, (b) $\tilde{\chi}_1^+\tilde{\chi}_1^- $ production with slepton/sneutrino-mediated-decays and (c) slepton pair production. In the model with intermediate sleptons, all three flavours ($\tilde{e}$, $\tilde{\mu}$, $\tilde{\tau}$) are included, while only $\tilde{e}$ and $\tilde{\mu}$ are included in the direct slepton model. In the final state, $\ell$ stands for an electron or muon, which can be produced directly or, in the case of (a) and (b) only, via a leptonically decaying $\tau$-lepton with additional neutrinos.}

\label{fig:c1c1viawwfeynman}
\end{figure}

The design of the analysis and the interpretation of results are based on simplified models~\cite{Alwall:2008ag}, where the masses of relevant sparticles (in this case the $\chinoonepm$, $\tilde{\ell}$, $\tilde{\nu}$ and $\ninoone$) are the only free parameters.
The $\tilde{\chi}_1^\pm$ is assumed to be pure wino and two possible decay modes are considered. The first is a decay into the $\tilde{\chi}_1^0$  via emission of a $W$ boson, which may decay into an electron or muon plus neutrino(s) either directly or through the emission of a leptonically decaying $\tau$-lepton (Figure \ref{fig:c1c1viawwfeynman}(a)). The second decay mode proceeds via a slepton--neutrino/sneutrino--lepton pair (Figure \ref{fig:c1c1viawwfeynman}(b)). In this case it is assumed that the scalar partners of the left-handed charged leptons and neutrinos are also light and thus accessible in the sparticle decay chains. It is also assumed they are mass-degenerate, and their masses are chosen to be midway between the mass of the chargino and that of the $\tilde{\chi}_1^0$, which is pure bino. Equal branching ratios for the three slepton flavours are assumed and charginos decay into charged sleptons or sneutrinos with a branching ratio of 50\% to each.
Lepton flavour is conserved in all models. In models with direct $\tilde\ell\tilde\ell$ production (Figure \ref{fig:c1c1viawwfeynman}(c)), each slepton decays into a lepton and a  $\tilde{\chi}_1^0$ with a 100\% branching ratio. Only $\tilde{e}$ and $\tilde{\mu}$ are considered in these models, and different assumptions about the masses of the superpartners of the left-handed and right-handed charged leptons, $\tilde{e}_\mathrm{L}$, $\tilde{e}_\mathrm{R}$, $\tilde{\mu}_\mathrm{L}$ and $\tilde{\mu}_\mathrm{R}$, 
are considered.

\section{ATLAS detector}
\label{sec:detector}
 
The ATLAS detector~\cite{PERF-2007-01} at the LHC is a general-purpose detector with a forward--backward symmetric cylindrical geometry and an almost complete coverage in solid angle around the collision point.\footnote{{
ATLAS uses a right-handed coordinate system with its origin at the nominal interaction point (IP) in the centre of the detector and the $z$-axis along the beam pipe.
The $x$-axis points from the IP to the centre of the LHC ring, and the $y$-axis points upwards.
Cylindrical coordinates $(r,\phi)$ are used in the transverse plane, $\phi$ being the azimuthal angle around the $z$-axis.
The pseudorapidity is defined in terms of the polar angle $\theta$ as $\eta = -\ln \tan(\theta/2)$. Rapidity is defined as $y=0.5\ln[(E+p_z)/(E-p_z)]$, where $E$ and $p_z$ denote the energy and the component of the particle momentum along the beam direction, respectively.
}}
It consists of an inner tracking detector surrounded by a thin superconducting solenoid, electromagnetic and hadronic calorimeters, and a muon spectrometer incorporating three large superconducting toroid magnets.
 
The inner-detector (ID) system is immersed in a \SI{2}{\tesla} axial magnetic field produced by the solenoid and provides charged-particle tracking in the range $|\eta| < 2.5$. It consists of a high-granularity silicon pixel detector, a silicon microstrip tracker and a transition radiation tracker, which enables radially extended track reconstruction up to $|\eta| = 2.0$.
The transition radiation tracker also provides electron identification information. 
During the first LHC long shutdown, a new tracking layer, known as the Insertable $B$-Layer~\cite{ATLAS-TDR-19,Abbott:2018ikt}, was added with an average sensor radius of 33 mm from the beam pipe to improve tracking and $b$-tagging performance.
 
The calorimeter system covers the pseudorapidity range $|\eta| < 4.9$.
Within the region $|\eta|< 3.2$, electromagnetic calorimetry is provided by barrel and endcap high-granularity lead/liquid-argon (LAr) sampling calorimeters.
Hadronic calorimetry is provided by an iron/scintillator-tile sampling calorimeter 
for $|\eta| < 1.7$, and two copper/LAr hadronic endcap calorimeters.
The solid angle coverage is completed with forward copper/LAr and tungsten/LAr calorimeter modules optimised for electromagnetic and hadronic measurements, respectively.
 
The muon spectrometer (MS) comprises separate trigger and high-precision tracking chambers measuring the deflection of muons in a magnetic field generated by superconducting air-core toroids.
The precision chamber system covers the region $|\eta| < 2.7$ with three layers of monitored drift tubes, complemented by cathode strip chambers in the forward region, where the background is higher.
The muon trigger system covers the range $|\eta| < 2.4$ with resistive plate chambers in the barrel, and thin gap chambers in the endcap regions.
 
A two-level trigger system is used to select events. There is a low-level hardware trigger implemented in custom electronics, which reduces the incoming data rate to a design value of 100 kHz using a subset of detector information, and a high-level software trigger that selects interesting final-state events with algorithms accessing the full detector information, and further reduces the rate to about 1 kHz~\cite{TRIG-2016-01}.

\section{Data and simulated event samples}
\label{sec:mcsamples}
 
The analysis uses data collected by the ATLAS detector during $pp$ collisions at a centre-of-mass energy of $\sqrt{s} = 13$~\TeV\ from 2015 to 2018.
The average number $\langle \mu \rangle$ of additional $pp$ interactions per bunch crossing (pile-up) ranged from 14 in 2015 to about 38 in 2017--2018.
After data-quality requirements, the data sample amounts to a total integrated luminosity of 139~fb$^{-1}$.
The uncertainty in the combined 2015--2018 integrated luminosity is 1.7\% \cite{ATLAS-CONF-2019-021}, obtained using the LUCID-2 detector \cite{LUCID2} for the primary luminosity measurements.
 
Candidate events were selected by a trigger that required at least two leptons (electrons and/or muons).
The trigger-level thresholds for the transverse momentum, \pT, of the leptons involved in the trigger decision were different according to the data-taking periods. They were in the range 8--22~\GeV\ for data collected in 2015 and 2016, and 8--24~\GeV\ for data collected in 2017 and 2018. These thresholds are looser than those applied in the lepton offline selection to ensure that trigger efficiencies are constant in the relevant phase space.
 
Simulated event samples are used for the SM background estimate and to model the SUSY signal.
The MC samples were processed through a full simulation of the ATLAS detector~\cite{SOFT-2010-01} based on \GEANT4~\cite{Agostinelli:2002hh} or a fast simulation using a parameterisation of the ATLAS calorimeter response and \GEANT4 for the other components of the detector~\cite{SOFT-2010-01}. They were reconstructed with the same algorithms as those used for the data. To compensate for differences 
between data and simulation in the lepton reconstruction efficiency, energy scale, energy resolution and modelling of the trigger~\cite{EGAM-2018-01,PERF-2015-10}, and in the $b$-tagging efficiency~\cite
{FTAG-2018-01}, correction factors are derived from data and applied as weights to the simulated events.
 
All SM backgrounds used are listed in Table \ref{tab:mc_backgr} along with the relevant parton distribution function (PDF) sets, the configuration of underlying-event and hadronisation parameters (tune), and the cross-section 
order in $\alpha_\mathrm{s}$ used to normalise the event yields for these samples. Further information on the ATLAS simulations of \ttbar, single top ($Wt$), multiboson and boson plus jet processes can be found in the relevant public notes~\cite{ATL-PHYS-PUB-2018-009,ATL-PHYS-PUB-2016-004,ATL-PHYS-PUB-2017-005,ATL-PHYS-PUB-2017-006}.
 
The SUSY signal samples were generated from leading-order (LO) matrix elements with up to two extra partons using \MGMCatNLO 2.6.1~\cite{Alwall:2014hca} interfaced to \PYTHIAV{8.186}~\cite{Sjostrand:2008}, with the A14 tune~\cite{ATL-PHYS-PUB-2014-021}, for the modelling of the SUSY decay chain, parton showering, hadronisation and the description of the underlying event. Parton luminosities were provided by the NNPDF2.3LO PDF set~\cite{Ball:2012cx}. Jet--parton matching was performed following the CKKW-L prescription~\cite{Lonnblad:2012ix}, with a matching scale set to one quarter of the mass of the pair-produced SUSY particles. Signal cross-sections were calculated to next-to-leading order (NLO) in 
$\alpha_\mathrm{s}$ adding the resummation of soft gluon emission at next-to-leading-logarithm accuracy (NLO+NLL)~\cite{Debove:2010kf,Fuks:2012qx,Fuks:2013vua,Fiaschi:2018hgm,Bozzi:2007qr,Fuks:2013lya,Fiaschi:2018xdm}. The nominal cross-sections and their uncertainties were taken from an envelope of cross-section predictions using different PDF sets and factorisation and renormalisation scales, as described in Ref.~\cite{Borschensky:2014cia}. The cross-section for $\tilde{\chi}_1^+\tilde{\chi}_1^- $ production, each with a mass of 400~\GeV, is $58.6\pm 4.7$~fb, while the cross-section for ${\tilde\ell\tilde\ell}$ production, each with a mass of 500~\GeV, is 
$0.47\pm 0.03$~fb for each generation of left-handed sleptons and 
$0.18\pm 0.01$~fb for each generation of right-handed sleptons.
 
Inelastic $pp$ interactions were generated and overlaid onto the hard-scattering process to simulate the effect of multiple proton--proton interactions occurring during the same (in-time) or a nearby (out-of-time) bunch crossing. These were produced using \PYTHIAV{8.186} and EvtGen~\cite{Lange:2001uf} with the NNPDF2.3LO set of
PDFs~\cite{Ball:2012cx} and the A3 tune~\cite{ATL-PHYS-PUB-2016-017}. The MC samples were reweighted so that the distribution of the average number of interactions per bunch crossing reproduces the observed distribution in the data.
\begin{DIFnomarkup}

\begin{table}[!h]
\tiny
\centering
\caption{\label{tab:mc_backgr}Simulated background event samples with the corresponding matrix element and parton shower (PS) generators, cross-section order in $\alpha_\mathrm{s}$ used to normalise the event yield, underlying-event tune and the generator PDF sets used.}
\begin{tabular}{lp{0.17\linewidth}p{0.08\linewidth}p{0.12\linewidth}p{0.11\linewidth}p{0.15\linewidth}p{0.12\linewidth}}
\toprule
 
Physics process          			& Generator 	& Parton shower 	&  Normalisation 		& Tune 			& PDF (generator) & PDF (PS)\\
\midrule
 
$t\bar{t}$ 				& \POWHEGBOXV{v2}~\cite{Frixione:2007,Frixione:2007vw,Alioli:2010,Campbell:2015}       		& \PYTHIAV{8.230}~\cite{Sjostrand:2014zea}  		& NNLO+NNLL~\cite{Czakon:2011xx} 		& A14~\cite{ATL-PHYS-PUB-2014-021} & NNPDF3.0NLO~\cite{Ball:2014uwa} & NNPDF2.3LO~\cite{Ball:2012cx}\\
$t\bar{t}+V (V=W,Z)$ 			& \MGMCatNLO~\cite{Alwall:2014hca}			& \PYTHIAV{8.210}~\cite{Sjostrand:2014zea}						& NLO~\cite{ATL-PHYS-PUB-2016-005,Alwall:2014hca} 			& A14 & NNPDF3.0NLO
& NNPDF2.3LO
\\
$t\bar{t}+WW$				& \MGMCatNLO			& \PYTHIAV{8.186}~\cite{Sjostrand:2008} 		& NLO~\cite{Alwall:2014hca}			& A14			& NNPDF2.3LO
&  NNPDF2.3LO
\\  
$tZ,t\bar{t}t\bar{t},t\bar{t}t$		& \MGMCatNLO			& \PYTHIAV{8.230}
& NLO~\cite{Alwall:2014hca}      		& A14			& NNPDF3.0NLO
& NNPDF2.3LO
\\
Single top ($Wt$) 				& \POWHEGBOXV{v2}~\cite{Frixione:2007vw,Alioli:2010,Re:2011}  	& \PYTHIAV{8.230}	&NLO+NNLL~\cite{Aliev:2010zk,Kant:2014oha}		& A14 			& NNPDF3.0NLO
& NNPDF2.3LO
\\
$Z/\gamma^*(\rightarrow ll )$+jets 	& \SHERPAV{2.2.1}~\cite{Gleisberg:2009,Bothmann:2019yzt,ATL-PHYS-PUB-2016-003}			& \SHERPAV{2.2.1}		& NNLO~\cite{Gavin:2010az}			& \SHERPA default~\cite{ATL-PHYS-PUB-2016-003}	& NNPDF3.0NNLO~\cite{Ball:2014uwa} &NNPDF3.0NNLO~\cite{Ball:2014uwa} \\
 
$WW,WZ,ZZ$				& \POWHEGBOXV{v2}~\cite{Frixione:2007vw,Alioli:2010,Melia:2011tj,Nason:2013ydw} 		& \PYTHIAV{8.210}
& NLO~\cite{ATL-PHYS-PUB-2017-005,Melia:2011tj,Nason:2013ydw}   		& AZNLO~\cite{STDM-2012-23}
&  CT10 NLO~\cite{Lai:2010vv} & CTEQ6L1\cite{Pumplin:2002} \\
$VVV (V=W,Z)$				& \SHERPAV{2.2.2}~\cite{Gleisberg:2009,Bothmann:2019yzt,ATL-PHYS-PUB-2017-005} 			&  \SHERPAV{2.2.2}		& NLO~\cite{Bothmann:2019yzt,ATL-PHYS-PUB-2017-005}			& \SHERPA default~\cite{ATL-PHYS-PUB-2017-005} 	& NNPDF3.0NNLO
& NNPDF3.0NNLO
\\
Higgs boson     				& \POWHEGBOXV{v2}~\cite{Frixione:2007vw,Alioli:2009,Alioli:2010,Nason:2010,Hamilton:2013,Hamilton:2015}			& \PYTHIAV{8.186}		       & NLO~\cite{deFlorian:2016spz}			& AZNLO			& NNPDF3.0NLO\footnote{The PDF4LHC15 set have been used for some Higgs production processes, as via gluon-gluon fusion, VBF and VH.} 
& CTEQ6L1
\\
\bottomrule
 
\end{tabular}
 
\end{table}

\end{DIFnomarkup}

\section{Object identification}
\label{sec:objevent}
Leptons selected for analysis are categorised as baseline or signal leptons according to various quality and kinematic selection criteria. Baseline objects are used in the calculation of missing transverse momentum, to resolve ambiguities between the analysis objects in the event and in the fake/non-prompt (FNP) lepton background estimation described in Section~\ref{sec:backgrounds}. Leptons used for the final event selection must satisfy more stringent signal requirements.
 
Baseline electron candidates are reconstructed using clusters of energy deposits in the electromagnetic calorimeter that are matched to an ID track. They are required to satisfy a \textit{Loose} likelihood-based identification requirement~\cite{EGAM-2018-01}, and to have $\pT>10$~\GeV\ and  $|\eta|<2.47$. They are also required to be within $|z_0 \sin\theta|$ = 0.5~mm of the primary vertex,\footnote{The primary vertex is defined as the vertex with the highest scalar sum of the squared transverse momentum of associated tracks with \pT$>500$~\MeV.} where $z_0$ is the longitudinal impact parameter relative to the primary vertex.
Signal electrons are required to satisfy a \textit{Tight} identification requirement~\cite{EGAM-2018-01} and the track associated with the signal electron is required to have $\vert d_0\vert/\sigma(d_0) < 5$, where $d_0$ is  the transverse impact parameter relative to the reconstructed primary vertex and $\sigma(d_0)$ is its error.
 
Baseline muon candidates are reconstructed in the pseudorapidity range $|\eta|<2.7$ from MS tracks matching ID tracks.
They are required to have $\pT>\SI{10}{\GeV}$, to be within $|z_0 \sin\theta|$ = 0.5~mm of the primary vertex and to satisfy the \textit{Medium} identification requirements defined in Ref.~\cite{PERF-2015-10}. The \textit{Medium} identification criterion defines requirements on the number of hits in the different ID and MS subsystems, and on the significance of the charge-to-momentum ratio $q/p$.
Finally, the track associated with the signal muon must have 
$\vert d_0\vert/\sigma(d_0) < 3$.
 
Isolation criteria are applied to signal electrons and muons. The scalar sum of the \pt of tracks inside a variable-size cone around the lepton (excluding its own track), must be less than 15\% of the lepton \pt. The track isolation cone size for electrons (muons) $\Delta R=\sqrt{(\Delta\eta)^2+(\Delta\phi)^2}$ is given by the minimum of $\Delta R = \SI{10}{~\GeV}/\pt$ and $\Delta R = 0.2\,(0.3)$.
In addition, for electrons (muons) the sum of the transverse energy of the calorimeter energy clusters in a cone of $\Delta R = 0.2$ around the lepton (excluding the energy from the lepton itself) must be less than 20\% (30\%) of the lepton \pt. For electrons with $\pT>200$ GeV these isolation requirements are not applied, and instead an upper limit of max($0.015\times\pT$, 3.5~\GeV) is placed on the transverse energy of the calorimeter energy clusters in a cone of $\Delta R = 0.2$ around the electron.
 
Jets are reconstructed from topological clusters of energy in the calorimeter~\cite{PERF-2014-07} using the anti-$k_t$ jet clustering algorithm~\cite{Cacciari:2008} as implemented in the FastJet package~\cite{Fastjet}, with a radius parameter $R=0.4$. The reconstructed jets are then calibrated by the application of a jet energy scale derived from 13~\TeV\ data and simulation~\cite{PERF-2016-04}.
Only jet candidates with $\pT>\SI{20}{\GeV}$ and $|\eta|<2.4$ are considered,\footnote{Hadronic $\tau$-lepton decay products are treated as jets.} although jets with $|\eta|<4.9$ are included in the missing transverse momentum calculation and are considered when applying the procedure to remove reconstruction ambiguities, which is described later in this Section.
 
To reduce the effects of pile-up, for jets with $|\eta|\leq 2.5$ and $\pt<\SI{120}{\GeV}$ a significant fraction of the tracks associated with each jet are required to have an origin compatible with the primary vertex, as defined by the jet vertex tagger~\cite{ATLAS-CONF-2014-018}. This requirement reduces jets from pile-up to 1\%, with an efficiency for pure hard-scatter jets of about 90\%. For jets with $|\eta|>2.5$ and $\pt<\SI{60}{\GeV}$, pile-up suppression is achieved through the forward jet vertex tagger~\cite{ATL-PHYS-PUB-2015-034}, which exploits topological correlations between jet pairs. Finally, events containing a jet that does not satisfy the jet-quality requirements~\cite{ATLAS-CONF-2015-029,PERF-2014-03} are rejected to remove events impacted by detector noise and non-collision backgrounds.
 
The MV2C10 boosted decision tree algorithm
~\cite{FTAG-2018-01} identifies jets containing $b$-hadrons (`$b$-jets') by using quantities such as the impact parameters of associated tracks, and well-reconstructed secondary vertices. A selection that provides 85\% efficiency for tagging $b$-jets in simulated $t\bar{t}$ events is used. The corresponding rejection factors against jets originating from $c$-quarks, from $\tau$-leptons, and from light quarks and gluons in the same sample at this working point are 2.7, 6.1 and 25, respectively.
 
To avoid the double counting of analysis baseline objects, a procedure to remove reconstruction ambiguities is applied as follows:
\begin{itemize}
\item jet candidates within $\Delta R' =\sqrt{\Delta y^2+\Delta\phi^2}$ = 0.2  of an electron candidate are removed; 
\item jets with fewer than three tracks that lie within $\Delta R' = 0.4$ of a muon candidate are removed;
\item electrons and muons within $\Delta R' = 0.4$ of the remaining jets are discarded, to reject leptons from the decay of $b$- or $c$-hadrons;
\item electron candidates 
are rejected if they are found to share an ID track with a muon.
\end{itemize}
 
The missing transverse momentum ($\mathbf p^\mathrm{miss}_\mathrm{T}$), which has the magnitude 
\MET, is defined as the negative vector sum of the transverse momenta of all identified physics objects (electrons, photons, muons and jets). Low-momentum tracks from the primary vertex that are not associated with reconstructed analysis objects (the `soft term') are also included in
the calculation, and the \MET value is adjusted for the calibration of the selected physics objects~\cite{PERF-2016-07}. Linked to the \MET value is the `object-based \MET significance', referred to as \MET significance in this paper, that helps to separate events with true \MET (arising from weakly interacting particles) from those where it is consistent with particle mismeasurement, resolution or identification inefficiencies. On an event-by-event basis, given the full event composition, \MET significance evaluates the p-value that the observed \MET is consistent with the null hypothesis of zero real \MET, as further detailed in Ref.~\cite{ATLAS-CONF-2018-038}.

\section{Search strategy}
\label{sec:SRs}
 
Events 
are required to have exactly two oppositely charged signal leptons $\ell_1$ and $\ell_2$, both with $\pT > 25~\GeV$. To remove contributions from low-mass resonances and to ensure good modelling of the SM background in all relevant regions, the invariant mass of the two leptons must be $m_{\ell_1\ell_2}>100$~\GeV. Events are further required to have no reconstructed $b$-jets, to suppress contributions from processes with top quarks. Selected events must also satisfy \MET$>110$~\GeV\ and \MET\ significance $>10$.
 
The stransverse mass $m_{\mathrm{T2}}$~\cite{Lester:1999tx,Barr:2003rg} is a kinematic variable used to
bound the masses of a pair of particles that are assumed to have each decayed into one visible and one invisible particle. It is defined as
\begin{equation*}
\mttwo
( \mathbf p_{\mathrm{T,1}}, \mathbf p_{\mathrm{T,2}},\mathbf p^\mathrm{miss}_\mathrm{T}) = \min_{\mathbf q_{\mathrm{T,1}} + \mathbf q_{\mathrm{T,2}} =
\mathbf p^\mathrm{miss}_\mathrm{T} }
\left\{ \max [\;
m_{\mathrm{T}}( \mathbf p_{\mathrm{T,1}}, \mathbf q_{\mathrm{T,1}} ),
m_{\mathrm{T}}( \mathbf p_{\mathrm{T,2}}, \mathbf q_{\mathrm{T,2}}
)
\;]
\right\},
\label{mt2eq}
\end{equation*}
 
\noindent where $m_{\mathrm T}$ indicates the transverse mass,\footnote{The transverse mass is defined as $m_\textup{T}=\sqrt{2\times|\mathbf p_{\mathrm{T,1}}|\times|\mathbf p_{\mathrm{T,2}}|\times(1-\mathrm{cos}(\Delta\phi))}$, where $\Delta\phi$ is the difference in azimuthal angle between the particles with transverse momenta $\mathbf p_{\mathrm{T,1}}$ and $\mathbf p_{\mathrm{T,2}}$.} 
$\mathbf p_{\mathrm{T,1}}$ and $\mathbf p_{\mathrm{T,2}}$ are the transverse-momentum vectors of the two leptons, and $\mathbf
q_{\mathrm{T,1}}$ and $\mathbf q_{\mathrm{T,2}}$ are vectors with
$\mathbf p^\mathrm{miss}_\mathrm{T}= \mathbf q_{\mathrm{T,1}} + \mathbf q_{\mathrm{T,2}}$.
The minimisation is performed over all the possible decompositions of $\mathbf p^\mathrm{miss}_\mathrm{T}$.
For $t\bar{t}$ or $WW$ decays, assuming an ideal detector with perfect momentum resolution,
$m_\textup{T2}( \mathbf p_{\mathrm{T,\ell_1}}, \mathbf p_{\mathrm{T,\ell_2}},\mathbf p^\mathrm{miss}_\mathrm{T})$
has a kinematic endpoint at the mass of the $W$~boson~\cite{Barr:2003rg}.
Signal models with significant mass splittings between the $\chinoonepm$ and the $\ninoone$ feature $\mttwo$ distributions that extend beyond the kinematic endpoint expected from the dominant SM backgrounds. Therefore, events are required to have high $m_{\mathrm{T2}}$ values.
 
Events are separated into `same flavour' (SF) events, i.e. $e^{\pm}e^{\mp}$ and $\mu^{\pm}\mu^{\mp}$, and `different flavour' (DF) events, i.e. $e^{\pm}\mu^{\mp}$, since the two classes of events have different background compositions. SF events are required to have a dilepton invariant mass far from the $Z$ peak,  $m_{\ell_1\ell_2} > 121.2$~\GeV,  to reduce diboson and $Z$+jets backgrounds.
 
Events are further classified by the multiplicity of non-$b$-tagged jets ($n_{\textrm{non-}b\textrm{-tagged jets}}$), i.e. the number of jets not identified as $b$-jets by the MV2C10 boosted decision tree algorithm. All events are required to have no more than one non-$b$-tagged jet.
Following the classification of the events, two sets of signal regions (SRs) are defined: a set of exclusive, `binned' SRs, to maximise model-dependent search sensitivity, and a set of `inclusive' SRs, to be used for model-independent results. Among the second set of SRs two are fully inclusive, with a different lower bound on $\mttwo$ to target different chargino or slepton mass regions, while two have both lower and upper bounds on $\mttwo$ to target models with lower endpoints. The definitions of these regions are shown in Table~\ref{tab:SRdefs}. Each SR is identified by the lepton flavour combination (DF or SF), the number of non-$b$-tagged jets (0J,1J) and the range of the $\mttwo$ interval.
 
\begin{DIFnomarkup}
 
\begin{table}[!htb]
\caption{\label{tab:SRdefs}
The definitions of the binned and inclusive signal regions. Relevant kinematic variables are defined in the text.
The bins labelled `DF' or `SF' refer to signal regions with different lepton flavour or same lepton flavour pair combinations, respectively, and the `0J' and `1J' labels refer to the multiplicity of non-$b$-tagged jets.
}
\begin{center}
\small
\begin{tabular}{l|cccc}
\hline
\multicolumn{1}{l}{} & & & &  \\[\dimexpr-\normalbaselineskip-\arrayrulewidth]
Signal region (SR)                       & SR-DF-0J & SR-DF-1J  & SR-SF-0J & SR-SF-1J \\
\hline
$n_{\text{non-} b \text{-tagged jets}}$                       & = 0 & = 1  & = 0  & = 1  \\
\hline
$m_{\ell_1\ell_2}$ [\GeV]             & \multicolumn{2}{c}{$>$100} & \multicolumn{2}{c}{$>$121.2} \\
\MET	[\GeV]	              & \multicolumn{4}{c}{$>$110} \\
\MET significance             & \multicolumn{4}{c}{$>$10} \\
$n_{b \text{-tagged jets}}$               & \multicolumn{4}{c}{= 0} \\
\hline
\multicolumn{5}{l}{Binned SRs} \\
\hline
 
\multirow{6}{*}{\mttwo\ [\GeV]} & \multicolumn{4}{c}{   \footnotesize $\in$[100,105) }  \\
& \multicolumn{4}{c}{\footnotesize $\in$[105,110) } \\
& \multicolumn{4}{c}{\footnotesize $\in$[110,120) }  \\
& \multicolumn{4}{c}{\footnotesize $\in$[120,140) }  \\
& \multicolumn{4}{c}{\footnotesize $\in$[140,160) }  \\
& \multicolumn{4}{c}{\footnotesize $\in$[160,180) }  \\
& \multicolumn{4}{c}{\footnotesize $\in$[180,220) }  \\
& \multicolumn{4}{c}{\footnotesize $\in$[220,260) }  \\
& \multicolumn{4}{c}{\footnotesize $\in$[260,$\infty$) }  \\
\hline
\multicolumn{5}{l}{Inclusive SRs} \\
\hline
\multirow{3}{*}{\mttwo\ [\GeV]} & \multicolumn{4}{c}{\footnotesize $\in$[100,$\infty$) }  \\
& \multicolumn{4}{c}{\footnotesize $\in$[160,$\infty$) }  \\
& \multicolumn{4}{c}{\footnotesize $\in$[100,120) }  \\
& \multicolumn{4}{c}{\footnotesize $\in$[120,160) }  \\
\hline
\end{tabular}
\end{center}
\end{table}
 
\end{DIFnomarkup}

\section{Background estimation and validation}
\label{sec:backgrounds}
 
The SM backgrounds can be classified into irreducible backgrounds, from processes with prompt leptons, 
and reducible backgrounds, which contain one or more FNP leptons. The main irreducible backgrounds come from SM diboson ($WW$, $WZ$, $ZZ$) and top-quark (\ttbar\ and $Wt$) production. These are estimated from simulated events, normalised using a simultaneous likelihood fit to data (as described in Section~\ref{sec:results}) in dedicated control regions (CRs). The CRs are designed to be enriched in the particular background process under study while remaining kinematically similar to the SRs. The normalisations of the relevant backgrounds are then validated in a set of validation regions (VRs), which are not used to constrain the fit, but are used to verify that the data and predictions, in terms of the yields and of the shapes of the relevant kinematic distributions, agree within uncertainties in regions of the phase space kinematically close to the SRs. Three CRs are used, as defined in Table~\ref{tab:CRdefs}:~\ CR-WW, targeting $WW$ production; CR-VZ, targeting $WZ$ and $ZZ$ production, which are normalised by using a single parameter in the likelihood fit to the data; and CR-top, targeting \ttbar\ and single-top-quark production, 
which are also normalised by using a single parameter
in the likelihood fit to the data. A single normalisation parameter is used for \ttbar\ and single-top-quark ($Wt$) production as the relative amounts of each process are consistent within uncertainties in the CR and SRs.
 
\begin{DIFnomarkup}
 
\begin{table}[!htb]
\caption{\label{tab:CRdefs}
Control region definitions for extracting normalisation factors for the dominant background processes. `DF' or `SF' refer to signal regions with different lepton flavour or same lepton flavour pair combinations, respectively.
}
\small
\begin{center}
\begin{tabular}{l|c|c|c}
\hline
Region                 					& CR-WW			& CR-VZ			& CR-top   \\
 
\hline
Lepton flavour                 	 			& DF				& SF				&  DF    \\
$n_{ b \text{-tagged jets}}$			& $= 0$ 			& $= 0$  			& $= 1$ \\
$n_{\text{non-} b \text{-tagged jets}}$ 	& $= 0$ 			& $= 0$         		& $= 0$ \\
 
$\mttwo$ [\GeV]		         			&$\in$ [60,65]		& $>120$ 			& $>80$ \\
$\MET$ [\GeV]	        					&$\in$ [60,100] 		& $>110$ 			& $>110$ \\
\MET significance					& $\in$ [5,10] 		& $>10$ 			& $>10$ \\
$m_{\ell_1\ell_2}$ [\GeV] 					& $>100$ 			& $\in$ [61.2,121.2]  & $>100$\\
\hline
\end{tabular}
\end{center}
\end{table}
 
\end{DIFnomarkup}

The definitions of the VRs are shown in Table~\ref{tab:VRdefs}. For the $WW$ background two validation regions are considered (VR-WW-0J and VR-WW-1J), according to the multiplicity of non-$b$-tagged jets in the event. As contributions from top-quark backgrounds in VR-WW-0J and VR-WW-1J are not negligible, three VRs are defined for this background. VR-top-low requires a similar \mttwo\ range as VR-WW-0J and VR-WW-1J, thus allowing the modelling of top-quark production at lower values of \mttwo\ to be validated. VR-top-high requires $\mttwo>100$~\GeV\ and provides validation in the high \mttwo\ region where the SRs are also defined.
Finally, VR-top-WW requires the same \MET, \MET\ significance and $\mttwo$ ranges as CR-WW and provides validation of the modelling of top-quark production in this region.
 
\begin{DIFnomarkup}
 
\begin{table}[!htb]
\caption{\label{tab:VRdefs}
Validation region definitions used to study the modelling of the SM backgrounds. `DF' or `SF' refer to regions with different lepton flavour or same lepton flavour pair combinations, respectively.
}
\small
\begin{center}
\begin{tabular}{l|c|c|c|c|c|c}
\hline
Region                    & VR-WW-0J      & VR-WW-1J                & VR-VZ                & VR-top-low							& VR-top-high	& VR-top-WW \\
\hline
Lepton flavour         	 & DF       & DF                 & SF        & DF					& DF & DF \\
$n_{ b \text{-tagged jets}}$	& $= 0$	& $= 0$                & $= 0$                &$= 1$ &$= 1$ &$= 1$ \\
$n_{\text{non-} b \text{-tagged jets}}$ 	 & $= 0$	& $= 1$            & $= 0$                & $= 0$						& $= 1$ & $= 1$\\
$\mttwo$ [\GeV]		& $\in$ [65,100]	&$\in$ [65,100] & $\in$ [100,120] & $\in$ [80,100] & $>100$	& $\in$ [60,65] \\
\met	[\GeV]		& $>60$	& $>60$			  & $>110$		    & $>110$						& $>110$	& $\in$ [60,100] \\
\met significance	& $>5$	& $>5$			  & $>10$		    & $\in$ [5,10]						& $>10$	&$\in$  [5,10] \\
$m_{\ell_1\ell_2}$ [\GeV]		 & $>100$       & $>100$                   & $\in$ [61.2,121.2]                   & $>100$					& $>100$	& $>100$ \\
\hline
\end{tabular}
\end{center}
\end{table}
\end{DIFnomarkup}
 
To obtain CRs and VRs of reasonable purity in $WW$ production,  CR-WW, VR-WW-0J and VR-WW-1J all require lower \mttwo\ values than the SRs. To validate the tails of the \mttwo\ distribution, a method similar to the one described in Ref.~\cite{CMS-SUS-17-010} is used. Three-lepton events, purely from $WZ$ production, are selected by requiring the absence of $b$-tagged jets and the presence of one same-flavour opposite-sign (SFOS) lepton pair with an invariant mass consistent with that of the $Z$~boson ($|m_{\ell_1\ell_2}-m_{Z}|<10$~\GeV). To avoid overlaps with portions of the phase space relevant for other searches, three-lepton events are also required to satisfy $\MET \in [40,170]$~\GeV. To emulate the signal regions, events are also required to have zero or one non-$b$-tagged jet. The transverse momentum of the lepton in the SFOS pair that has the same charge as the remaining lepton is added to the $\mathbf p^\mathrm{miss}_\mathrm{T}$ vector, to mimic a neutrino. The \mttwo\ value can then be calculated using the remaining two leptons in the event. With this selection, there is a good agreement between the shapes of the \mttwo\ distributions  observed in data and simulation, and no additional systematic uncertainty is applied to the $WW$ background at high \mttwo.
 
Sub-dominant irreducible SM background contributions come from Drell--Yan, \ttbar+$V$ and Higgs boson production. These processes, jointly referred to as `Other backgrounds'  (or `Others' in the Figures) are estimated directly from simulation using the samples described in Section~\ref{sec:mcsamples}.
The remaining background from FNP leptons is estimated from data using the matrix method (MM)~\cite{TOPQ-2010-01}. This method considers two types of lepton identification criteria: `signal' leptons, corresponding to leptons passing the full analysis selection, and `baseline' leptons, as defined in Section~\ref{sec:objevent}.
Probabilities for prompt leptons satisfying the baseline selection to also satisfy the signal selection are measured as a function of lepton \pt\ and $\eta$ in dedicated regions enriched in $Z$~boson processes.
Similar probabilities for FNP leptons are measured in events dominated by leptons from the decays of heavy-flavour hadrons and from photon conversions.
These probabilities are used in the MM to extract data-driven estimates for the FNP lepton background in the CRs, VRs, and SRs, comparing the numbers of events containing a pair of baseline leptons in which one of the two leptons, both or none of them satisfy the signal selection in a given region. To avoid double counting between the simulated samples used for background estimation and the FNP lepton background estimate provided by the MM, all simulated events containing one or more FNP leptons are removed.
 
\begin{figure}[!htb]
\centering
\begin{subfigure}[b]{0.48\linewidth}
\includegraphics[width=\linewidth]{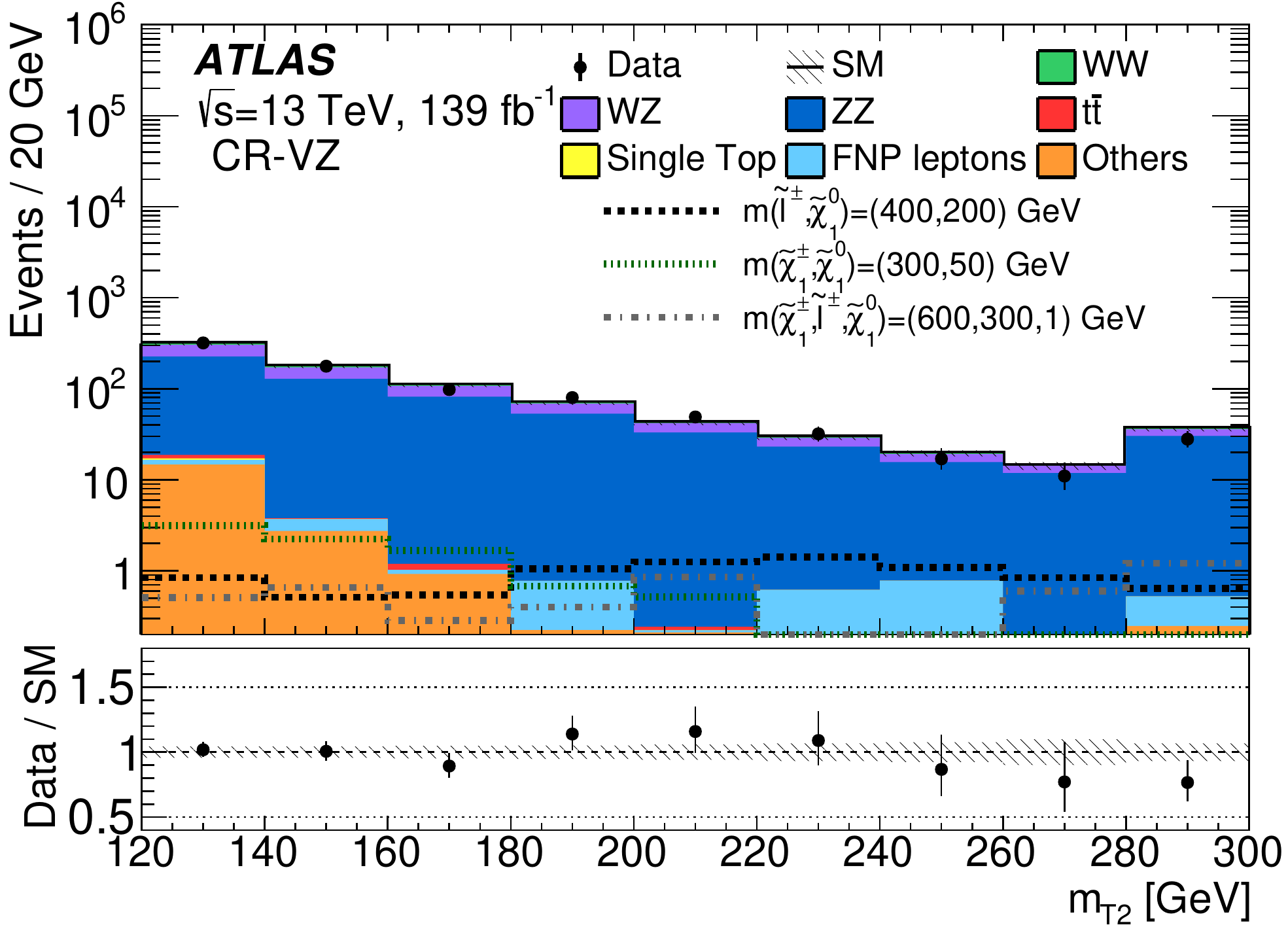}
\caption{\mttwo\ distribution in CR-VZ}
\end{subfigure}
\begin{subfigure}[b]{0.48\linewidth}
\includegraphics[width=\linewidth]{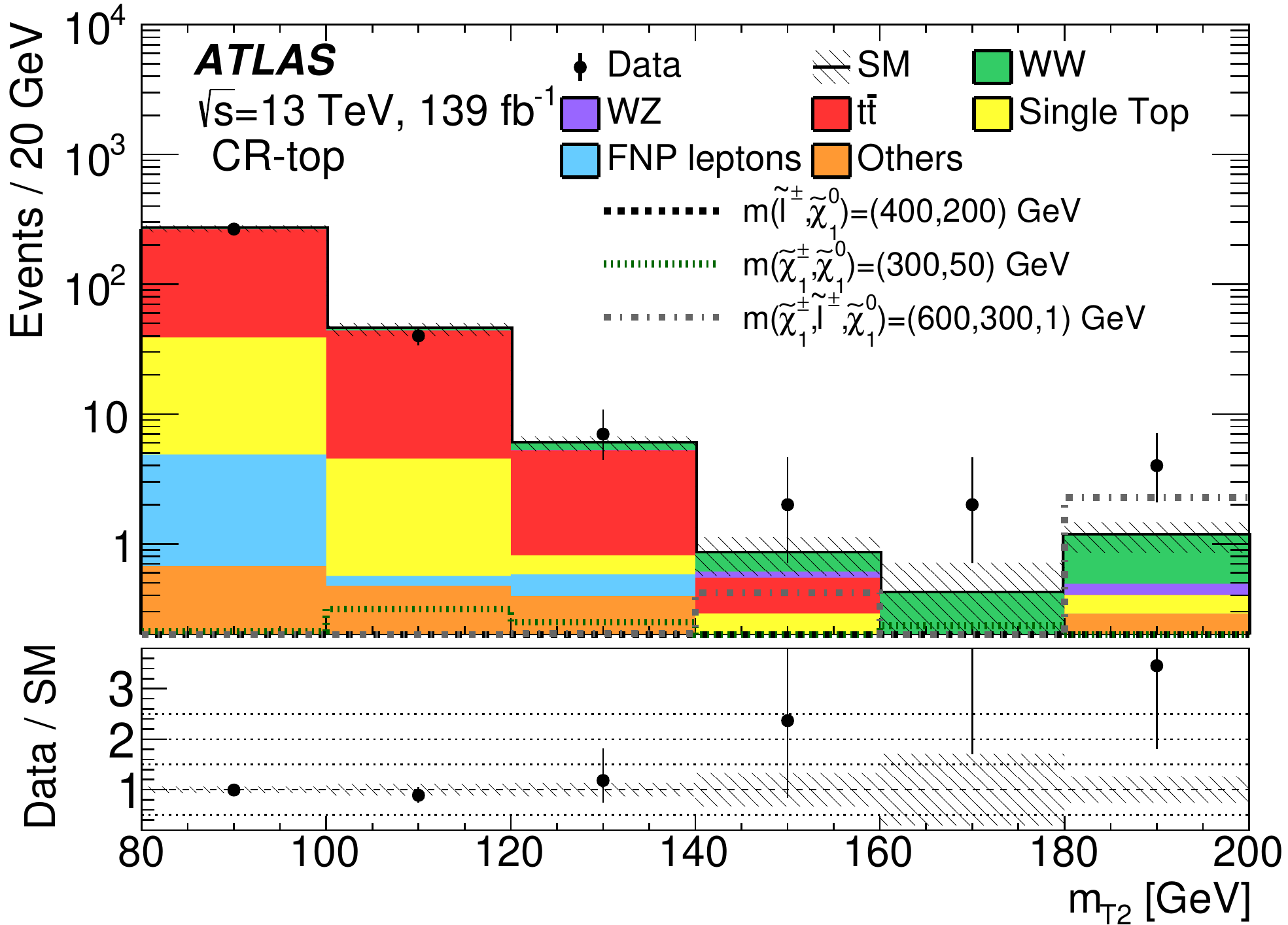}
\caption{\mttwo\ distribution in CR-top}
\end{subfigure}
\begin{subfigure}[b]{0.48\linewidth}
\includegraphics[width=\linewidth]{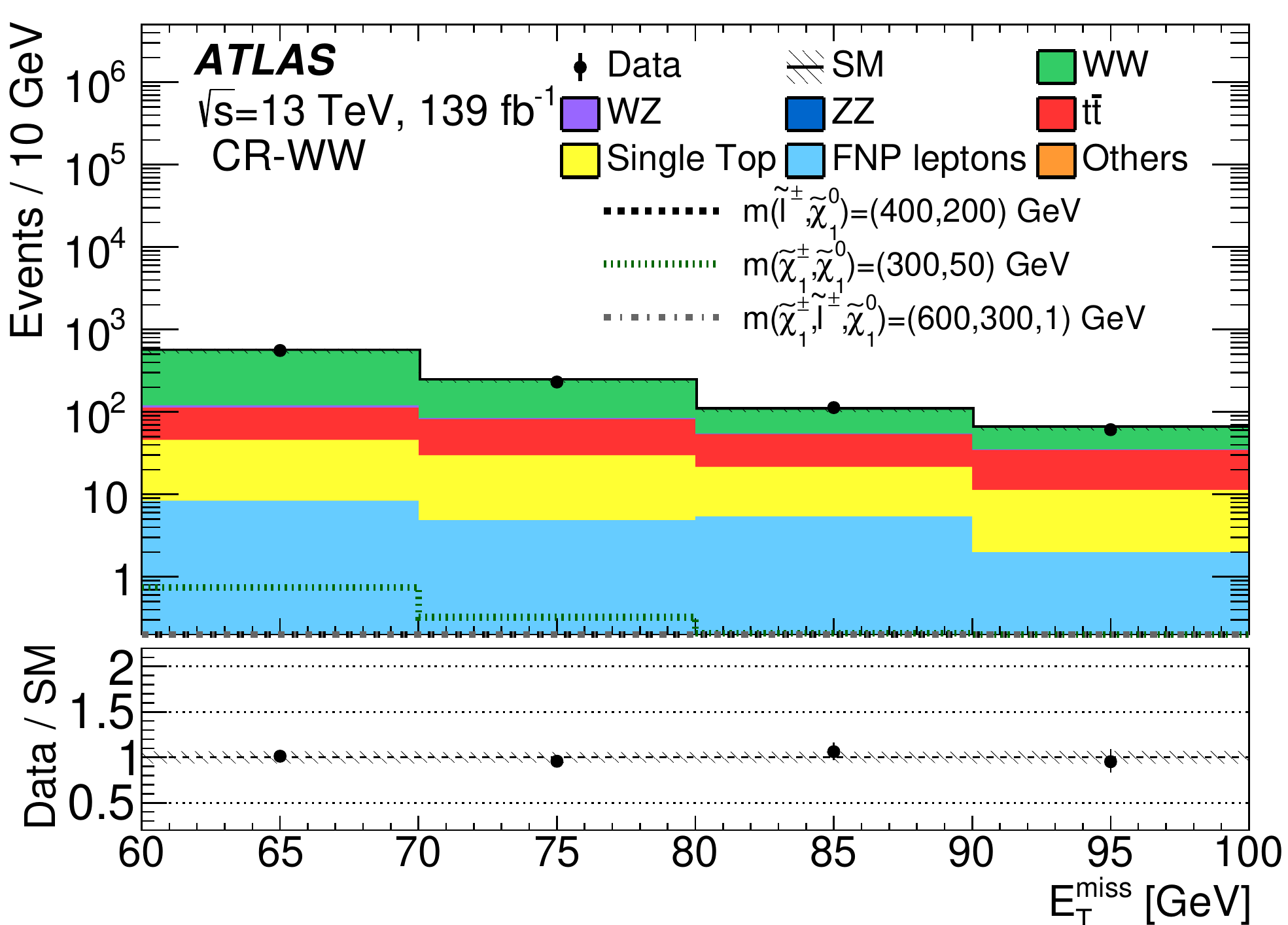}
\caption{\MET\ distribution in CR-WW}
\end{subfigure}
\caption{ Distributions of \mttwo\ in (a) CR-VZ and (b) CR-top and (c) \MET\ in CR-WW for data and the estimated SM backgrounds. The normalisation factors extracted from the corresponding CRs are used to rescale the $t\bar{t}$, single-top-quark, $WW$, $WZ$ and $ZZ$ backgrounds. The FNP lepton background is calculated using the data-driven matrix method. Negligible background contributions are not included in the legends. The uncertainty band includes systematic and statistical errors from all sources and the final bin in each histogram includes the overflow. Distributions for three benchmark signal points are overlaid for comparison. The lower panels show the ratio of data to the SM background estimate.}
\label{fig:CR}
\end{figure}

The number of observed events in each CR, as well as the predicted yield of each SM process, is shown in Table~\ref{tab:CRresults}.
For backgrounds whose normalisation is extracted from the likelihood fit, the yield expected from the simulation before the fit is also shown.
After the fit, the central value of the total number of predicted events in each CR matches the data, as expected from the normalisation procedure. The normalisation factors returned by the fit for the $WW$, $t\bar{t}$ and single-top-quark backgrounds, and $WZ$/$ZZ$ backgrounds are $1.25 \pm 0.11$, $0.82 \pm 0.06$ and $1.18 \pm 0.05$ respectively, which for diboson backgrounds are applied to MC samples scaled to NLO cross-sections (as detailed in Table \ref{tab:mc_backgr}).
The shapes of kinematic distributions are well reproduced by the simulation in each CR\@. The distributions of $\mttwo$ in CR-VZ and CR-top and of $\MET$ in CR-WW are shown in Figure~\ref{fig:CR}.
 
\begin{figure}[b]
\centering
\begin{subfigure}[b]{0.48\linewidth}
\includegraphics[width=\linewidth]{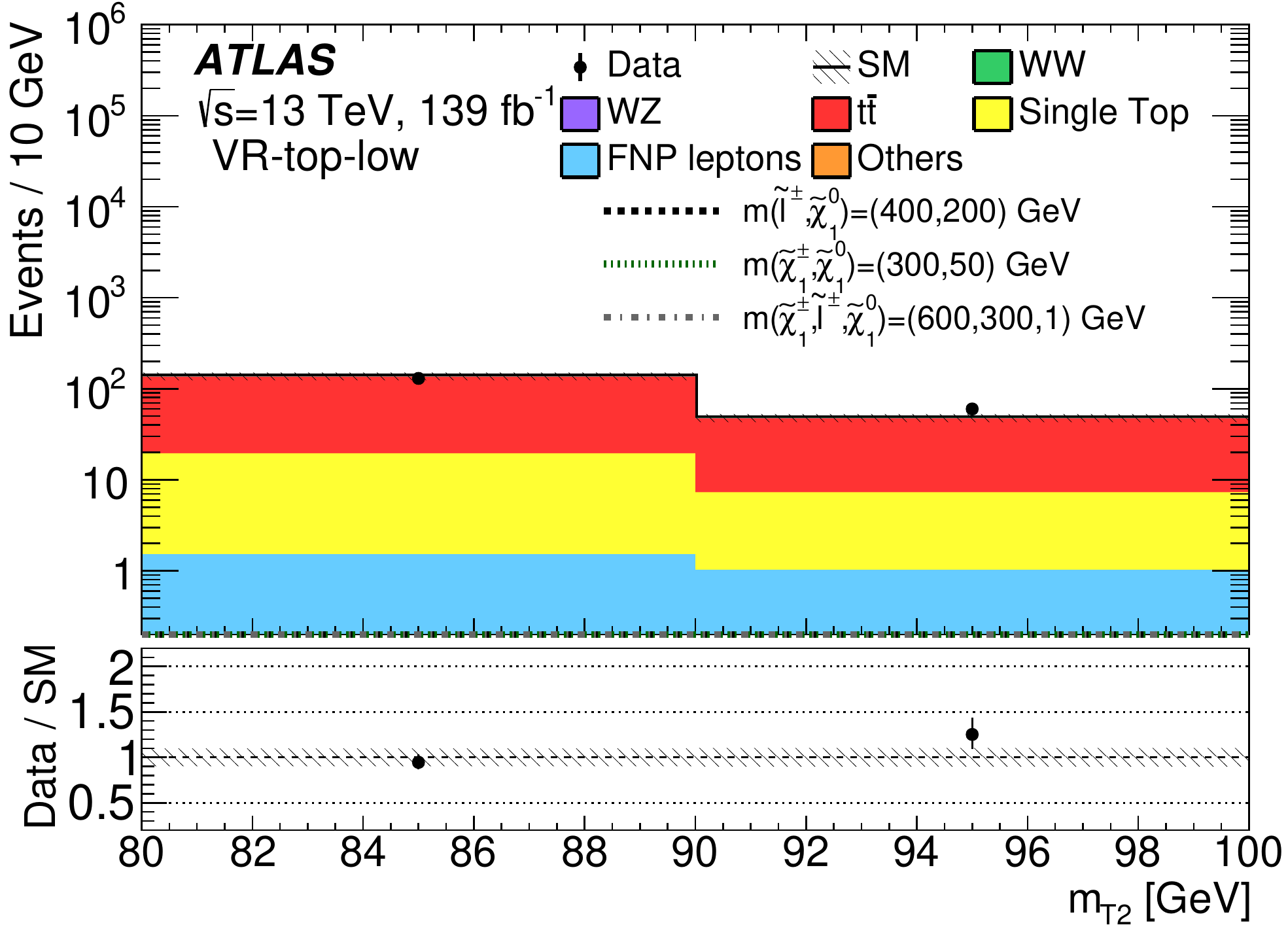}
\caption{\mttwo\ distribution in VR-top-low}
\end{subfigure}
\begin{subfigure}[b]{0.48\linewidth}
\includegraphics[width=\linewidth]{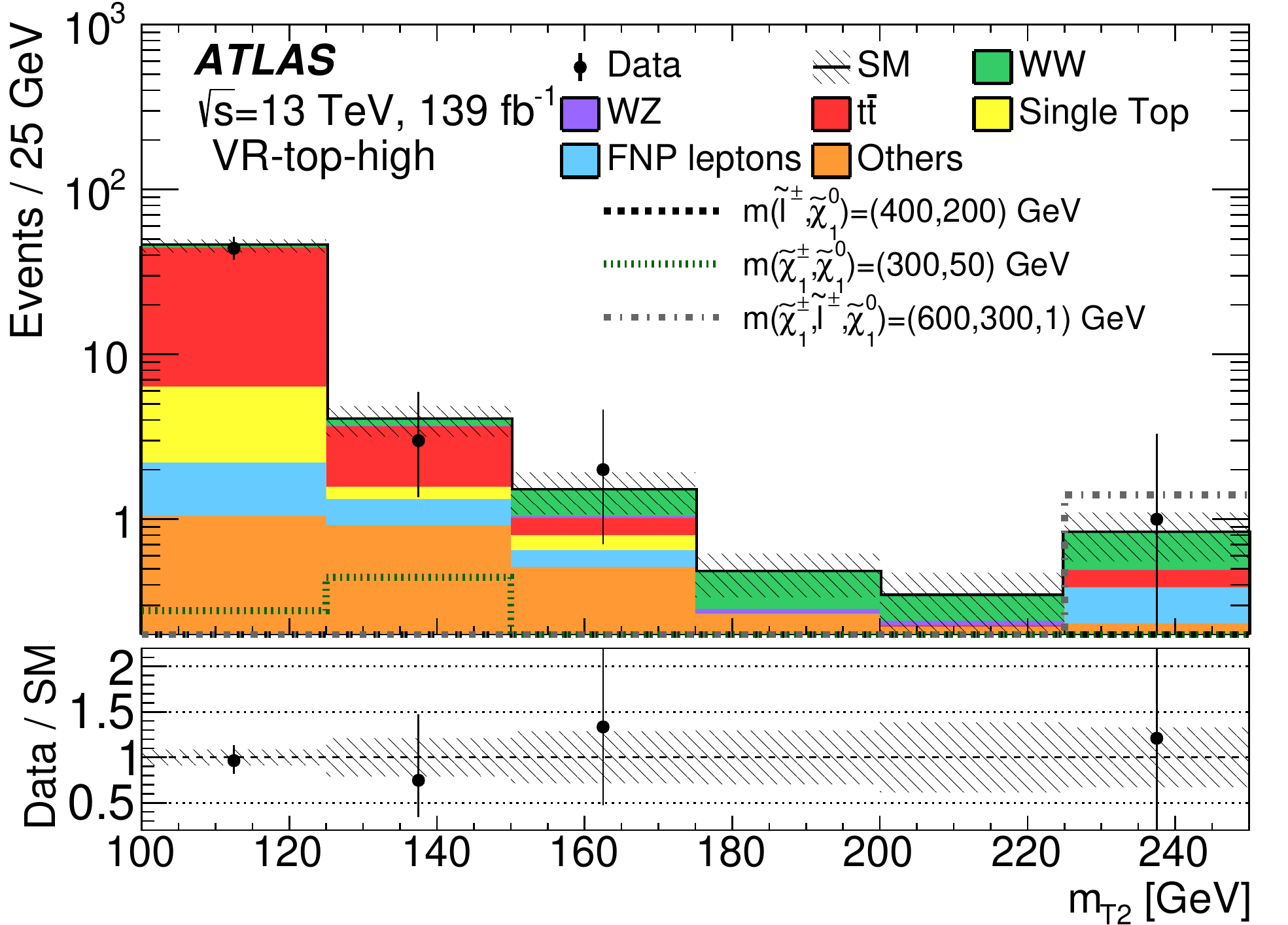}
\caption{\mttwo\ distribution in VR-top-high}
\end{subfigure}
\begin{subfigure}[b]{0.48\linewidth}
\includegraphics[width=\textwidth]{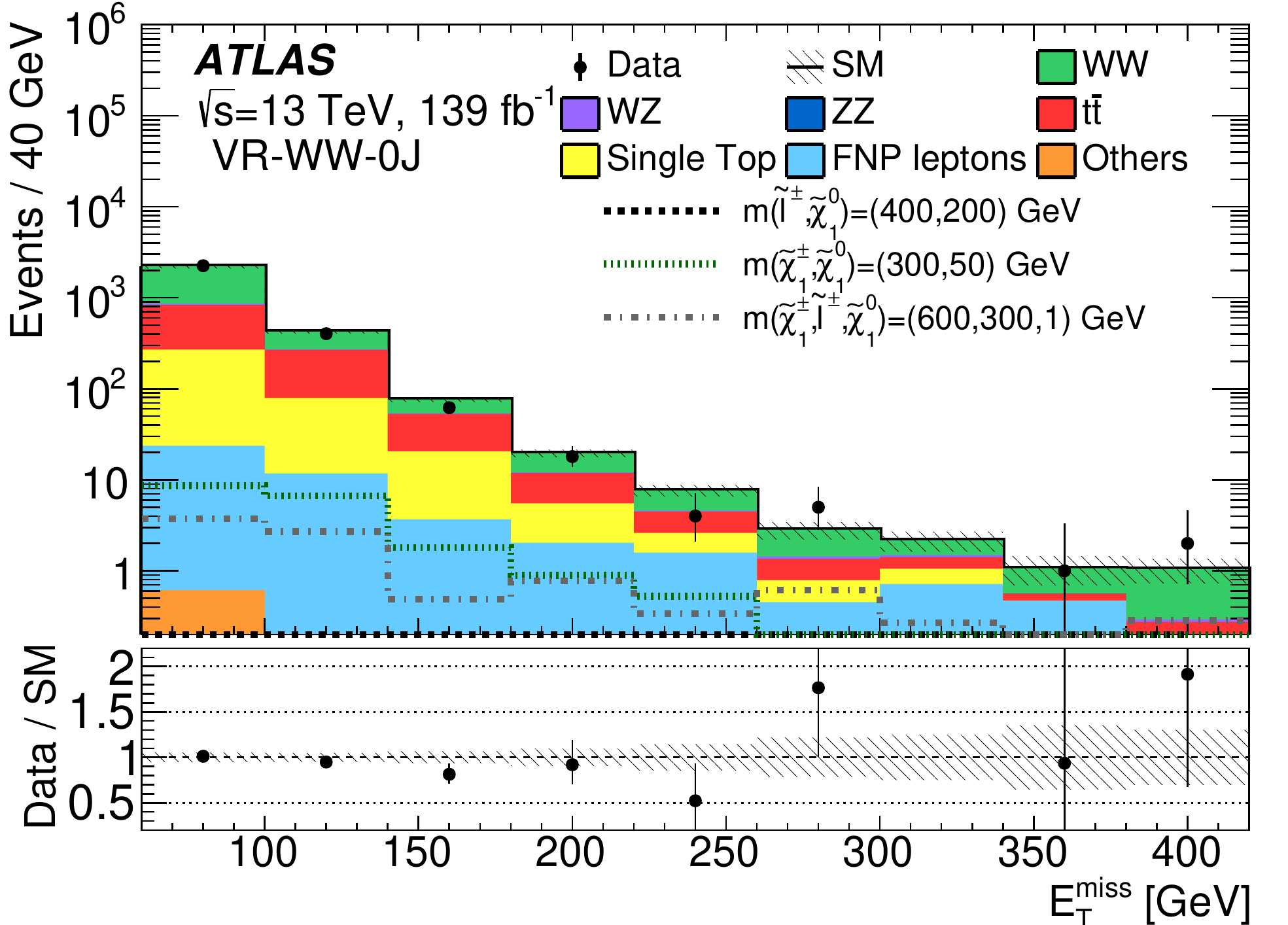}
\caption{\MET\ distribution in VR-WW-0J}
\end{subfigure}
\begin{subfigure}[b]{0.48\linewidth}
\includegraphics[width=\linewidth]{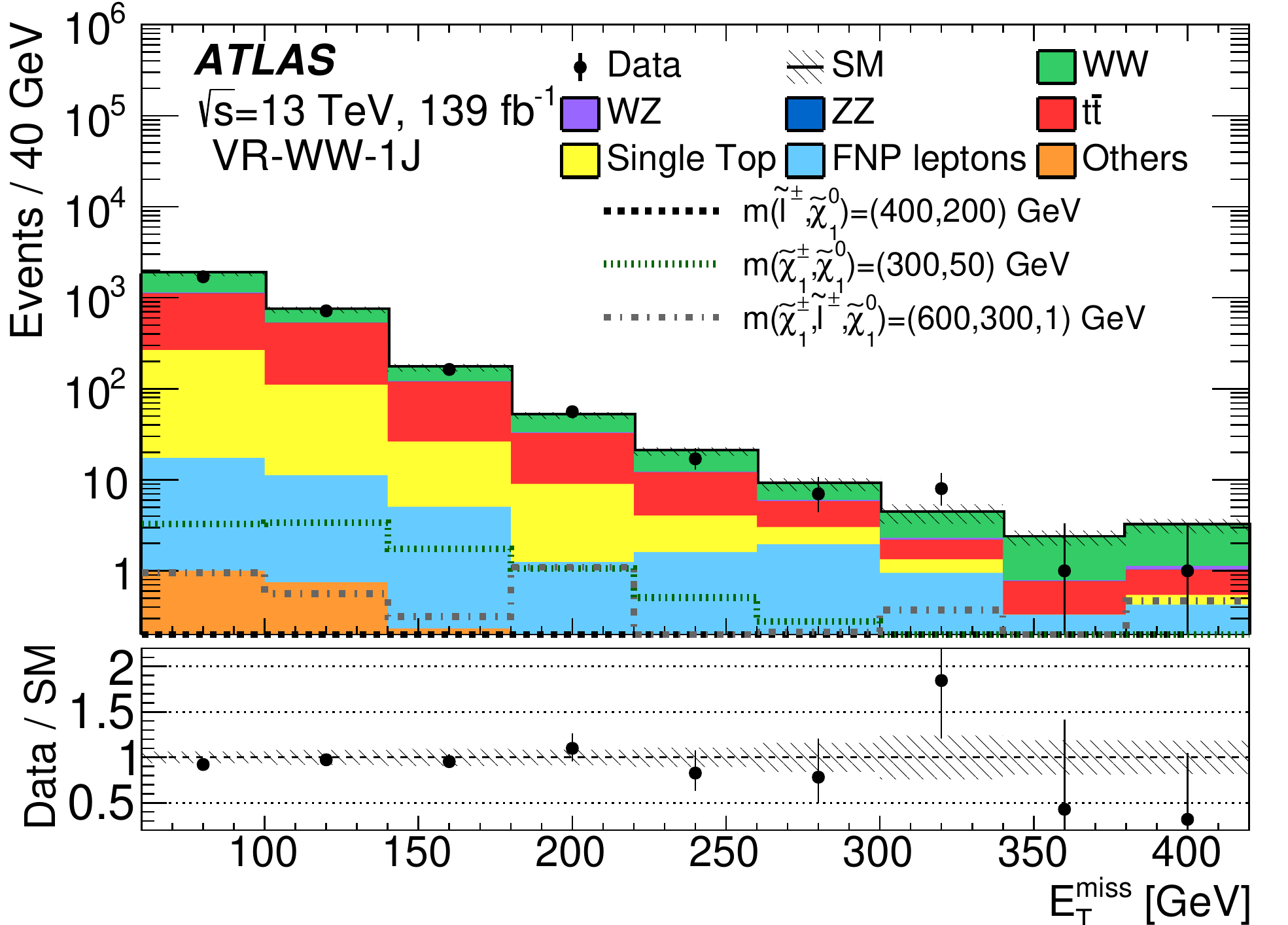}
\caption{\MET\ distribution in VR-WW-1J}
\end{subfigure}
\begin{subfigure}[b]{0.48\linewidth}
\includegraphics[width=\linewidth]{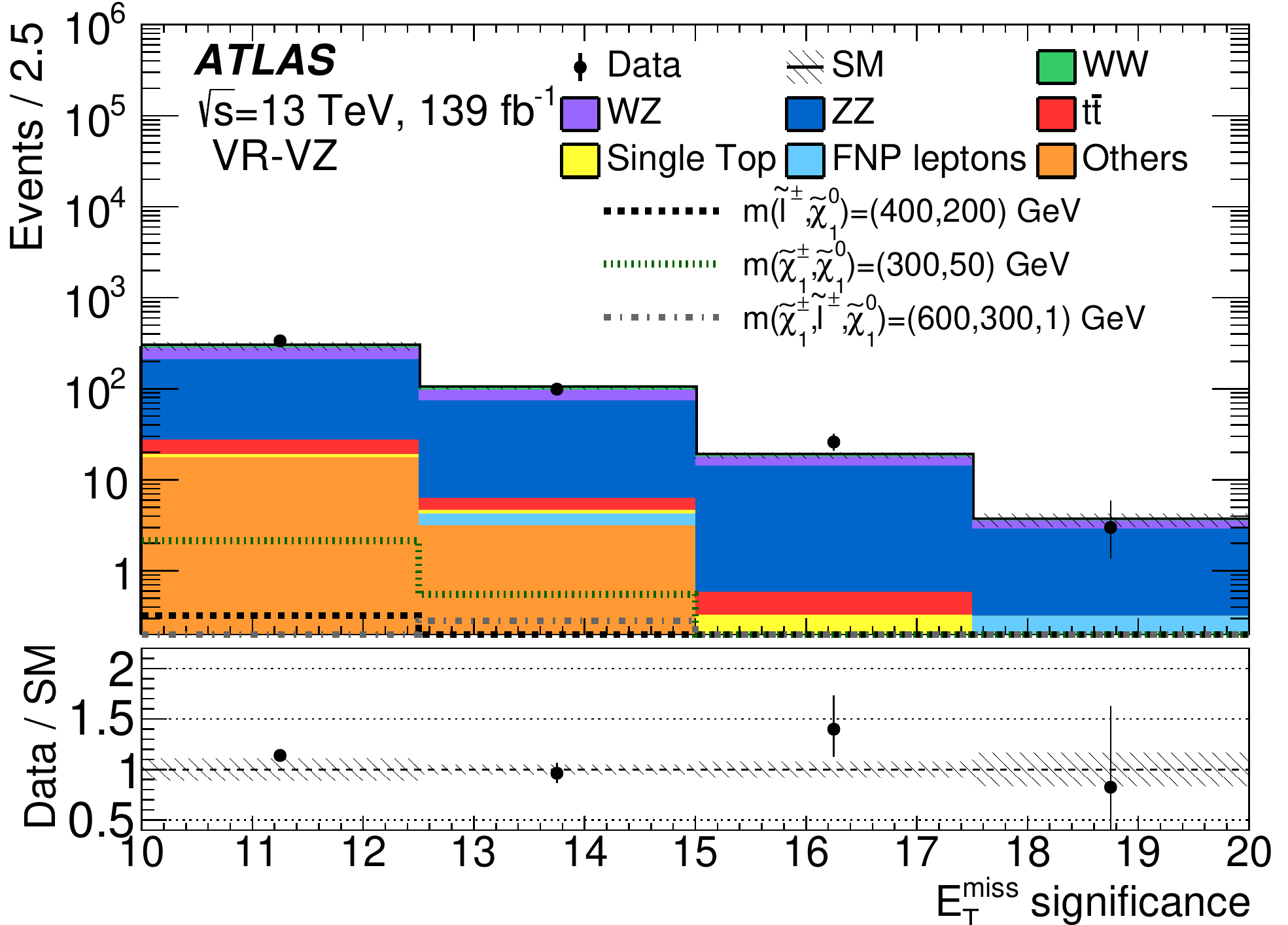}
\caption{\MET\ significance distribution in VR-VZ}
\end{subfigure}
\begin{subfigure}[b]{0.48\linewidth}
\includegraphics[width=\linewidth]{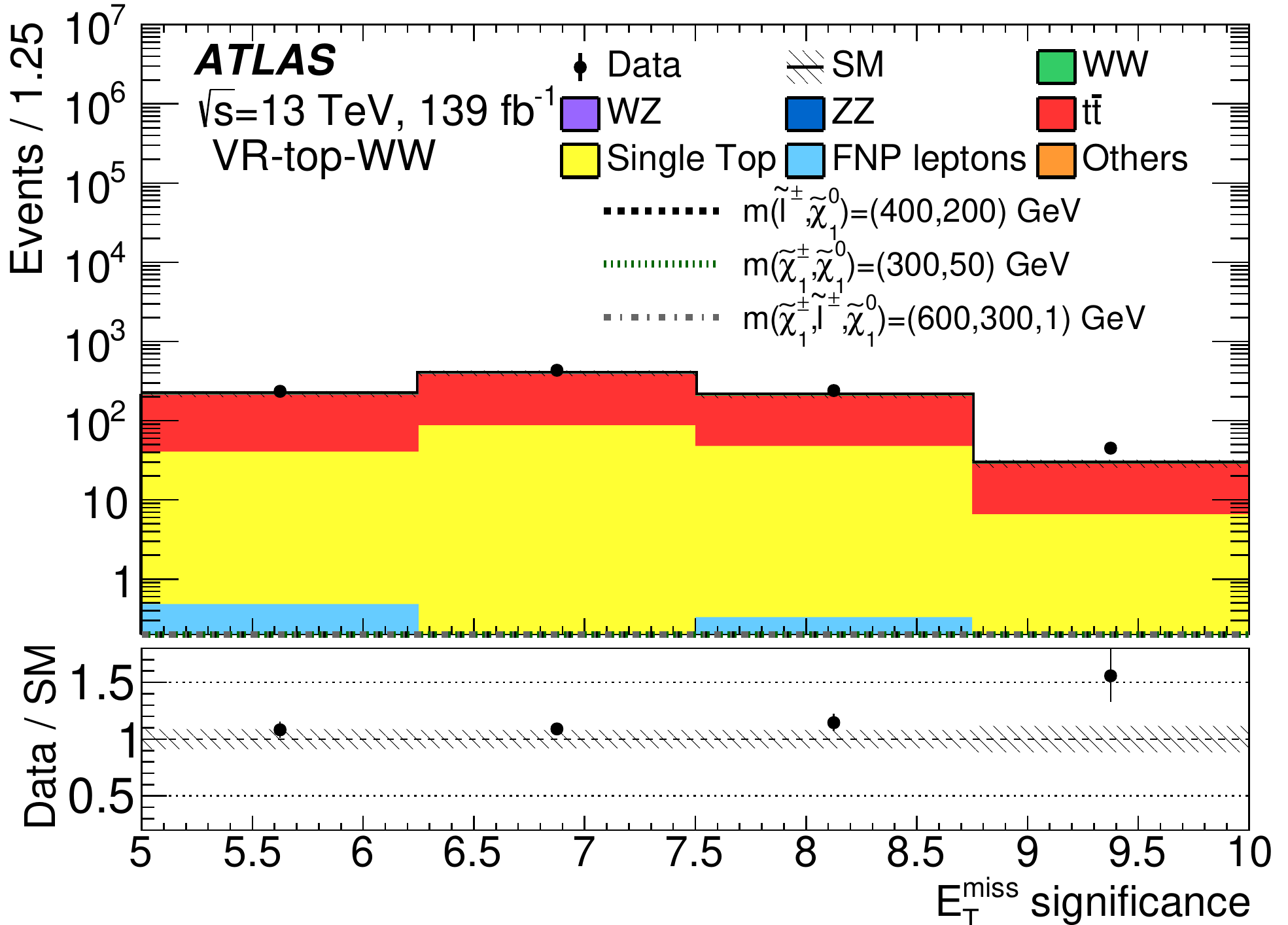}
\caption{\MET\ significance distribution in VR-top-WW}
\end{subfigure}
 
\caption{ Distributions of \mttwo\ in (a) VR-top-low  and (b) VR-top-high, \MET\ in (c) VR-WW-0J and (d) VR-WW-1J, and \MET\ significance in (e) VR-VZ and (f) VR-top-WW, for data and the estimated SM backgrounds. The normalisation factors extracted from the corresponding CRs are used to rescale the $t\bar{t}$, single-top-quark, $WW$, $WZ$ and $ZZ$ backgrounds. The FNP lepton background is calculated using the data-driven matrix method. Negligible background contributions are not included in the legends. The uncertainty band includes systematic and statistical errors from all sources and the last bin includes the overflow. Distributions for three benchmark signal points are overlaid for comparison. The lower panels show the ratio of data to the SM background estimate.
}
\label{fig:VRdistributions}
\end{figure}

The number of observed events and the predicted background in each VR are shown in Table~\ref{tab:VRresults}. For backgrounds with a normalisation extracted from the fit, the expected yield from simulated samples before the fit is also shown. Figure~\ref{fig:VRdistributions} shows a selection of kinematic distributions for data and the estimated SM background in the validation regions defined in Table~\ref{tab:VRdefs}. Good agreement is observed in all regions.
 
\begin{DIFnomarkup}

\begin{table}
\caption{
Observed event yields and predicted background yields from the fit in the CRs. For backgrounds with a normalisation extracted from the fit, the yield expected from the simulation before the fit is also shown. `Other backgrounds' include the non-dominant background sources, i.e. $t\bar{t}$+$V$, Higgs boson and Drell--Yan events. A `--' symbol indicates that the background contribution is negligible.
}
\begin{center}
\setlength{\tabcolsep}{0.0pc}
{\small
\begin{tabular*}{\textwidth}{@{\extracolsep{\fill}}lccc}
\noalign{\smallskip}\hline\noalign{\smallskip}
Region                  & CR-WW                                 & CR-VZ                         & CR-top                        \\[-0.05cm]
\noalign{\smallskip}\hline\noalign{\smallskip}
Observed events         & $962$                                 & $811$                         & $321$                         \\
\noalign{\smallskip}\hline\noalign{\smallskip}
Fitted backgrounds      & $962 \pm 31$ \hspace{4.5pt}           & $811 \pm 28$ \hspace{4pt}     & $321 \pm 18$  \hspace{4pt}    \\
\noalign{\smallskip}\hline\noalign{\smallskip}
{$\!\begin{aligned}
& \text{Fitted } WW         \\[-2pt]
& \text{Fitted } WZ         \\[-2pt]
& \text{Fitted } ZZ         \\[-2pt]
& \text{Fitted } t \bar{t}  \\[-2pt]
& \text{Fitted single top}  \\[-2pt]
& \text{Other backgrounds}  \\[-2pt]
& \text{FNP leptons}
\end{aligned}
$}
&   {$\!\begin{aligned}
670     &\pm 60     \\[-2pt]
11.8    &\pm 0.7    \\[-2pt]
0.29    &\pm 0.06   \\[-2pt]
170     &\pm 50     \\[-2pt]
88      &\pm 8      \\[-2pt]
0.17    &\pm 0.06   \\[-2pt]
21      &\pm 8
\end{aligned}
$}
&   {$\!\begin{aligned}
19.1   &\pm 1.9     \\[-2pt]
188    &\pm 7       \\[-2pt]
577    &\pm 23      \\[-2pt]
1.8    &\pm 1.3     \\[-2pt]
0.65   &\pm 0.35    \\[-2pt]
19     &\pm 7       \\[-2pt]
5&^{+6}_{-5}
\end{aligned}
$}
&   {$\!\begin{aligned}
5.5     &\pm 2.7    \\[-2pt]
0.32    &\pm 0.15   \\[-2pt]
&-                  \\[-2pt]
270     &\pm 16     \\[-2pt]
38.6    &\pm 2.6    \\[-2pt]
2.21    &\pm 0.20   \\[-2pt]
4.2     &\pm 2.2
\end{aligned}
$} \\
\noalign{\smallskip}\hline\noalign{\smallskip}
{$\!\begin{aligned}
& \text{Simulated } WW         \\[-2pt]
& \text{Simulated } WZ         \\[-2pt]
& \text{Simulated } ZZ         \\[-2pt]
& \text{Simulated } t \bar{t}  \\[-2pt]
& \text{Simulated single top}
\end{aligned}
$}
&   {$\!\begin{aligned}
528&     \\[-2pt]
9&.9     \\[-2pt]
0&.24    \\[-2pt]
210&     \\[-2pt]
107&
\end{aligned}
$}
&   {$\!\begin{aligned}
15&.1    \\[-2pt]
158&     \\[-2pt]
487&     \\[-2pt]
2&.2     \\[-2pt]
0&.8
\end{aligned}
$}
&   {$\!\begin{aligned}
4&.3     \\[-2pt]
0&.27    \\[-2pt]
&-       \\[-2pt]
327&     \\[-2pt]
46&.7
\end{aligned}
$} \\
\hline
\end{tabular*}
}
\end{center}
\label{tab:CRresults}
\end{table}
\end{DIFnomarkup}

\begin{DIFnomarkup}

\begin{table}
\caption{
Observed event yields and predicted background yields in the VRs. For backgrounds with a normalisation extracted from the fit in the CRs, the yield expected from the simulation before the fit is also shown. `Other backgrounds' include the non-dominant background sources, i.e. $t \bar{t}$+$V$, Higgs boson and Drell--Yan events. A `--' symbol indicates that the background contribution is negligible.
}
\begin{center}
\setlength{\tabcolsep}{0.0pc}
{\small
\begin{tabular*}{\textwidth}{@{\extracolsep{\fill}}lccccccc}
\noalign{\smallskip}\hline\noalign{\smallskip}
Regions                 & VR-WW-0J          & VR-WW-1J          & VR-VZ                 & VR-top-low            & VR-top-high           & VR-top-WW                 \\[-0.05cm]
\noalign{\smallskip}\hline\noalign{\smallskip}
Observed events         &   $2742$            & $2671$          & $464$                 & $190$                 & $50$                  & $953$                     \\
\noalign{\smallskip}\hline\noalign{\smallskip}
Fitted backgrounds      &   $2760 \pm 120$ \hspace{0.1pt} & $2840 \pm 250$ \hspace{0.001pt}   & $420 \pm 40$ \hspace{0.1pt}       & $185 \pm 17$ \hspace{4.5pt}         & $53 \pm 7$ \hspace{9.5pt}           & $850 \pm 80$ \hspace{4.5pt}             \\
\noalign{\smallskip}\hline\noalign{\smallskip}
{$\!\begin{aligned}
& \text{Fitted } WW         \\[-2pt]
& \text{Fitted } WZ         \\[-2pt]
& \text{Fitted } ZZ         \\[-2pt]
& \text{Fitted } t \bar{t}  \\[-2pt]
& \text{Fitted single top}  \\[-2pt]
& \text{Other backgrounds}  \\[-2pt]
& \text{FNP leptons}
\end{aligned}
$}
&   {$\!\begin{aligned}
1550    &\pm 150    \\[-2pt]
34.2    &\pm 2.0    \\[-2pt]
0.50    &\pm 0.06   \\[-2pt]
790     &\pm 110    \\[-2pt]
336     &\pm 32     \\[-2pt]
0.92    &\pm 0.30   \\[-2pt]
44      &\pm 23
\end{aligned}
$}
&   {$\!\begin{aligned}
990     &\pm 120    \\[-2pt]
27.0    &\pm 2.3    \\[-2pt]
0.39    &\pm 0.07   \\[-2pt]
1400    &\pm 270    \\[-2pt]
380     &\pm 40     \\[-2pt]
2.1     &\pm 0.5    \\[-2pt]
38      &\pm 21
\end{aligned}
$}
&   {$\!\begin{aligned}
17.6    &\pm 2.2    \\[-2pt]
99      &\pm 9      \\[-2pt]
268     &\pm 25     \\[-2pt]
10.5    &\pm 3.2    \\[-2pt]
2.2     &\pm 1.4    \\[-2pt]
21&^{+27}_{-21}     \\[-2pt]
0.2&_{-0.2}^{+2.1}
\end{aligned}
$}
&   {$\!\begin{aligned}
2.1     &\pm 0.7        \\[-2pt]
0.05&_{-0.05}^{+0.17}   \\[-2pt]
&-                      \\[-2pt]
157     &\pm 15         \\[-2pt]
24.3    &\pm 2.6        \\[-2pt]
0.28    &\pm 0.06       \\[-2pt]
2.3     &\pm 1.4
\end{aligned}
$}
&   {$\!\begin{aligned}
2.6     &\pm 1.4    \\[-2pt]
0.2&_{-0.2}^{+0.6}  \\[-2pt]
&-                  \\[-2pt]
40      &\pm 7      \\[-2pt]
4.6     &\pm 1.4    \\[-2pt]
3.20    &\pm 0.20   \\[-2pt]
1.8     &\pm 0.5
\end{aligned}
$}
&   {$\!\begin{aligned}
16.1    &\pm 2.5        \\[-2pt]
0.53    &\pm 0.13       \\[-2pt]
0.01&_{-0.01}^{+0.03}   \\[-2pt]
650     &\pm 70         \\[-2pt]
182     &\pm 15         \\[-2pt]
0.39    &\pm 0.11       \\[-2pt]
&-
\end{aligned}
$} \\
\noalign{\smallskip}\hline\noalign{\smallskip}
{$\!\begin{aligned}
& \text{Simulated } WW         \\[-2pt]
& \text{Simulated } WZ         \\[-2pt]
& \text{Simulated } ZZ         \\[-2pt]
& \text{Simulated } t \bar{t}  \\[-2pt]
& \text{Simulated single top}
\end{aligned}
$}
&   {$\!\begin{aligned}
1230&    \\[-2pt]
28&.8    \\[-2pt]
0&.42    \\[-2pt]
960&     \\[-2pt]
406&
\end{aligned}
$}
&   {$\!\begin{aligned}
790&      \\[-2pt]
22&.8     \\[-2pt]
0&.33     \\[-2pt]
1700&     \\[-2pt]
462&
\end{aligned}
$}
&   {$\!\begin{aligned}
14&.0      \\[-2pt]
84&        \\[-2pt]
226&       \\[-2pt]
13&        \\[-2pt]
2&.6
\end{aligned}
$}
&   {$\!\begin{aligned}
1&.6        \\[-2pt]
0&.04       \\[-2pt]
&-          \\[-2pt]
190&        \\[-2pt]
29&.4
\end{aligned}
$}
&   {$\!\begin{aligned}
2&.0         \\[-2pt]
0&.1         \\[-2pt]
&-          \\[-2pt]
49&          \\[-2pt]
5&.6
\end{aligned}
$}
&   {$\!\begin{aligned}
12&.8       \\[-2pt]
0&.45       \\[-2pt]
0&.01       \\[-2pt]
790&        \\[-2pt]
220&
\end{aligned}
$} \\
\noalign{\smallskip}\hline\noalign{\smallskip}
\end{tabular*}
}
\label{tab:VRresults}
\end{center}
\end{table}
 
\end{DIFnomarkup}

\section{Systematic uncertainties}
\label{sec:systematics}
 
All relevant sources of experimental and theoretical systematic uncertainty affecting the SM background estimates and the signal predictions are included in the likelihood fit described in Section~\ref{sec:results}. The dominant sources of systematic uncertainty are related to theoretical uncertainties in the MC modelling, while the largest sources of experimental uncertainty are related to the jet energy scale (JES) and jet energy resolution (JER). 
The statistical uncertainty in the simulated event samples is also accounted for. Since the normalisation of the predictions for the dominant background processes is extracted from dedicated control regions, the systematic uncertainties only affect the extrapolation to the signal regions in these cases.
 
The JES and JER uncertainties are considered as a function of jet \pT and $\eta$, the pile-up conditions and the  flavour composition of the selected jet sample. They are derived using a combination of data and simulation, through measurements of the transverse momentum balance between a jet and a reference object in dijet, $Z$+jets and $\gamma+$jets events~\cite{PERF-2016-04}. 
An additional uncertainty in $\mathbf p^\mathrm{miss}_\mathrm{T}$ comes from the soft-term resolution and scale
~\cite{PERF-2016-07}. Uncertainties in the scale factors applied to the simulated samples to account for differences between data and simulation in the $b$-jet identification efficiency are also included.
The remaining experimental systematic uncertainties, such as those in the lepton reconstruction efficiency, lepton energy scale and lepton energy resolution and differences between the trigger efficiencies in data and simulation are included and are found to be a few per mille
in all channels. The reweighting procedure (pile-up reweighting) applied to simulation to match the distribution of the number of
interactions per bunch crossing observed in data results in a negligible contribution to the total systematic uncertainty.

Several sources of theoretical uncertainty in the modelling of the dominant backgrounds are considered. Uncertainties in the MC modelling of diboson events are estimated by varying the PDF sets as well as the renormalisation and factorisation scales used to generate the samples. To account for effects due to the choice of generator, the nominal \POWHEGBOX diboson samples are compared with \SHERPA diboson samples that have a different matrix element calculation and parton shower simulation.
 
For \ttbar production, uncertainties in the parton shower simulation are estimated by comparing samples generated with \POWHEGBOX interfaced to either \PYTHIAV{8.186} or \HERWIG 7.04~\cite{Bahr:2008pv,Bellm:2015jjp}. Another source of uncertainty comes from the modelling of initial- and final-state radiation, which is calculated by comparing the predictions of the nominal sample with two alternative samples generated with \POWHEGBOX interfaced to \PYTHIAV{8.186} but with the radiation settings varied~\cite{ATL-PHYS-PUB-2016-020}. The uncertainty associated with the choice of event generator is estimated by comparing the nominal samples with samples generated with \AMCatNLO interfaced to \PYTHIAV{8.186}~\cite{ATL-PHYS-PUB-2017-007}. Finally, for single-top-quark production an uncertainty is assigned to the treatment of the interference between the $Wt$ and $t\bar{t}$ samples. This is done by comparing the nominal sample generated using the diagram removal method with a sample generated using the diagram subtraction method~\cite{ATL-PHYS-PUB-2016-020}.
 
There are several contributions to the uncertainty in the MM estimate of the FNP background. 
First, an uncertainty is included to account for the difference between the probability in simulation and the probability in data that prompt leptons may satisfy the signal selection.
Furthermore, uncertainties in the expected composition of the FNP leptons in the signal regions are included. Finally, two uncertainties associated with the control regions used to derive the probabilities for baseline leptons to satisfy the signal requirements are considered. The first accounts for limited numbers of events in these regions and the second for the subtraction of prompt-lepton contamination.

Systematic uncertainties on the signal acceptance and shape due to scale and parton shower variations are found to be negligible. The systematic uncertainty on the signal cross section has been described in Section~\ref{sec:mcsamples}.
 
A summary of the impact of the systematic uncertainties on the background yields in the inclusive SRs with $\mttwo>100$~\GeV, after performing the likelihood fit, is shown in Table~\ref{tab:systematicSummary}. For the binned SRs defined in Table~\ref{tab:SRdefs}, the impact of the uncertainties associated with the limited numbers of MC events is higher than for the inclusive SRs. 
 
\begin{DIFnomarkup}

\begin{table}
\caption{Summary of the dominant systematic uncertainties in the background estimates in the inclusive SRs requiring $\mttwo>100$~\GeV\ after performing the profile likelihood fit. The individual uncertainties can be correlated, and do not necessarily add in quadrature to
the total background uncertainty. The percentages show the size of the uncertainty relative to the total expected background. `Top theoretical uncertainties' refers to \ttbar theoretical uncertainties and the uncertainty associated with $Wt$--$\ttbar$ interference added in quadrature.
\label{tab:systematicSummary}}
\begin{center}
\setlength{\tabcolsep}{0.0pc}
\begin{tabular}{p{0.4\linewidth}p{0.16\linewidth}p{0.16\linewidth}p{0.16\linewidth}p{0.12\linewidth}}
\noalign{\smallskip}\hline\noalign{\smallskip}
{Region}            & \hspace{-10pt} SR-DF-0J               & \hspace{-10pt} SR-DF-1J               & \hspace{-10pt} SR-SF-0J               & \hspace{-10pt} SR-SF-1J            \\
$\mttwo$ [\GeV]     & \hspace{-10pt} $\in$[100,$\infty$)    & \hspace{-10pt} $\in$[100,$\infty$)    & \hspace{-10pt} $\in$[100,$\infty$)    & \hspace{-10pt} $\in$[100,$\infty$) \\
\noalign{\smallskip}\hline\noalign{\smallskip}
 
\end{tabular}
\begin{tabular}{p{0.4\linewidth}p{0.16\linewidth}p{0.16\linewidth}p{0.16\linewidth}p{0.12\linewidth}}
Total background expectation        & \hspace{1.7pt} 96    & \hspace{1.7pt} 75      & 144     & 124 \\
\noalign{\smallskip}\hline\noalign{\smallskip}
{$\!\begin{aligned}
& \text{MC statistical uncertainties }      \\[-2pt]
& WW \text{ normalisation}                  \\[-2pt]
& VZ \text{ normalisation}                  \\[-2pt]
& t\bar{t} \text{ normalisation}            \\[-2pt]
& \text{Diboson theoretical uncertainties}  \\[-2pt]
& \text{Top theoretical uncertainties}      \\[-2pt]
& \MET \text{ modelling}                    \\[-2pt]
& \text{Jet energy scale}                   \\[-2pt]
& \text{Jet energy resolution}              \\[-2pt]
& \text{Pile-up reweighting}                \\[-2pt]
& b\text{-tagging}                          \\[-2pt]
& \text{Lepton modelling}                   \\[-2pt]
& \text{FNP leptons}
\end{aligned}
$}
&   {$\!\begin{aligned}
3     &\%   \\[-2pt]
7     &\%   \\[-2pt]
<1    &\%   \\[-2pt]
1     &\%   \\[-2pt]
7     &\%   \\[-2pt]
7     &\%   \\[-2pt]
1     &\%   \\[-2pt]
2     &\%   \\[-2pt]
1     &\%   \\[-2pt]
<1    &\%   \\[-2pt]
<1    &\%   \\[-2pt]
1     &\%   \\[-2pt]
1     &\%
\end{aligned}
$}
&   {$\!\begin{aligned}
3     &\%   \\[-2pt]
6     &\%   \\[-2pt]
<1    &\%   \\[-2pt]
2     &\%   \\[-2pt]
7     &\%   \\[-2pt]
8     &\%   \\[-2pt]
1     &\%   \\[-2pt]
3     &\%   \\[-2pt]
2     &\%   \\[-2pt]
1     &\%   \\[-2pt]
2     &\%   \\[-2pt]
1     &\%   \\[-2pt]
1     &\%
\end{aligned}
$}
&   {$\!\begin{aligned}
2     &\%   \\[-2pt]
4     &\%   \\[-2pt]
1     &\%   \\[-2pt]
<1    &\%   \\[-2pt]
4     &\%   \\[-2pt]
3     &\%   \\[-2pt]
<1    &\%   \\[-2pt]
2     &\%   \\[-2pt]
1     &\%   \\[-2pt]
<1    &\%   \\[-2pt]
<1    &\%   \\[-2pt]
1     &\%   \\[-2pt]
1     &\%
\end{aligned}
$}
&   {$\!\begin{aligned}
3     &\%   \\[-2pt]
3     &\%   \\[-2pt]
1     &\%   \\[-2pt]
1     &\%   \\[-2pt]
3     &\%   \\[-2pt]
6     &\%   \\[-2pt]
2     &\%   \\[-2pt]
2     &\%   \\[-2pt]
2     &\%   \\[-2pt]
<1    &\%   \\[-2pt]
1     &\%   \\[-2pt]
3     &\%   \\[-2pt]
1     &\%
\end{aligned}
$} \\
\noalign{\smallskip}\hline\noalign{\smallskip}
{$\!\begin{aligned}
& \text{Total systematic uncertainties}
\end{aligned}
$}
&   {$\!\begin{aligned}
\hspace{5pt} 15 &\%
\end{aligned}
$}
&   {$\!\begin{aligned}
\hspace{5pt} 12 &\%
\end{aligned}
$}
&   {$\!\begin{aligned}
\hspace{10pt} 8 &\%
\end{aligned}
$}
&   {$\!\begin{aligned}
\hspace{5pt} 10 &\%
\end{aligned}
$} \\
\noalign{\smallskip}\hline\noalign{\smallskip}
\end{tabular}
\end{center}
\end{table}

\end{DIFnomarkup}

\section{Results}
\label{sec:results}
The statistical interpretation of the final results is performed using the HistFitter framework~\cite{HistFitter}. A simultaneous likelihood fit is performed, which includes either just the CRs (in the case of the background-only fit) or the CRs and one or more of the SRs (when calculating exclusion limits).  The likelihood is a product of Poisson probability density functions describing the observed number of events in each CR/SR and Gaussian distributions that constrain the nuisance parameters associated with the systematic uncertainties. Systematic uncertainties that are correlated between different samples are accounted for in the fit configuration by using the same nuisance parameter. These include the diboson theory uncertainties, for which a combined nuisance parameter is used for the WW, WZ and ZZ backgrounds.
The uncertainties are applied in each of the CRs and SRs and their effect  is correlated for events across all regions in the fit. Poisson distributions are used for MC statistical uncertainties.
 
\begin{figure}[htb!]
\centering
\includegraphics[width=\linewidth]{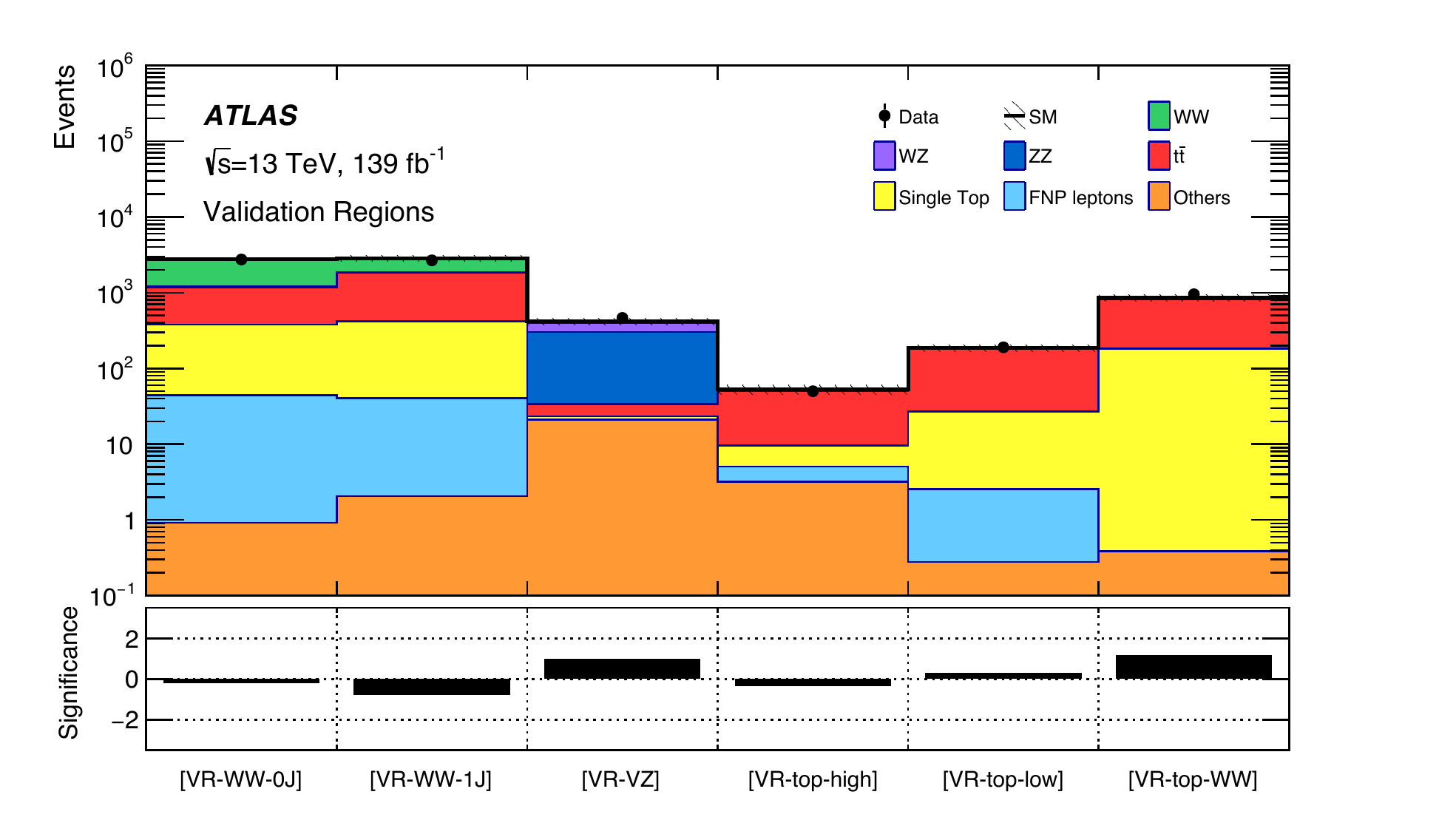}
\caption{The upper panel shows the observed number of events in each of the VRs defined in Table~\ref{tab:VRdefs}, together with the expected SM backgrounds obtained after the background-only fit in the CRs. The shaded band represents the total uncertainty in the expected SM background.  The lower panel shows the significance as defined in Ref.~\cite{Cousin:2008clt}.  
}
\label{fig:VRpull}
\end{figure}

A background-only fit that uses data only in the CRs is performed to constrain the nuisance parameters of the likelihood function, which include the background normalisation factors and parameters associated with the systematic uncertainties. The results of the background-only fit are used to assess how well the data agree with the background estimates in the validation regions. Good agreement, within about one standard deviation for all VRs, is observed, as described in Section~\ref{sec:backgrounds} and shown in Figure~\ref{fig:VRpull}.
 
\begin{figure}[h]
\centering
\includegraphics[width=\linewidth]{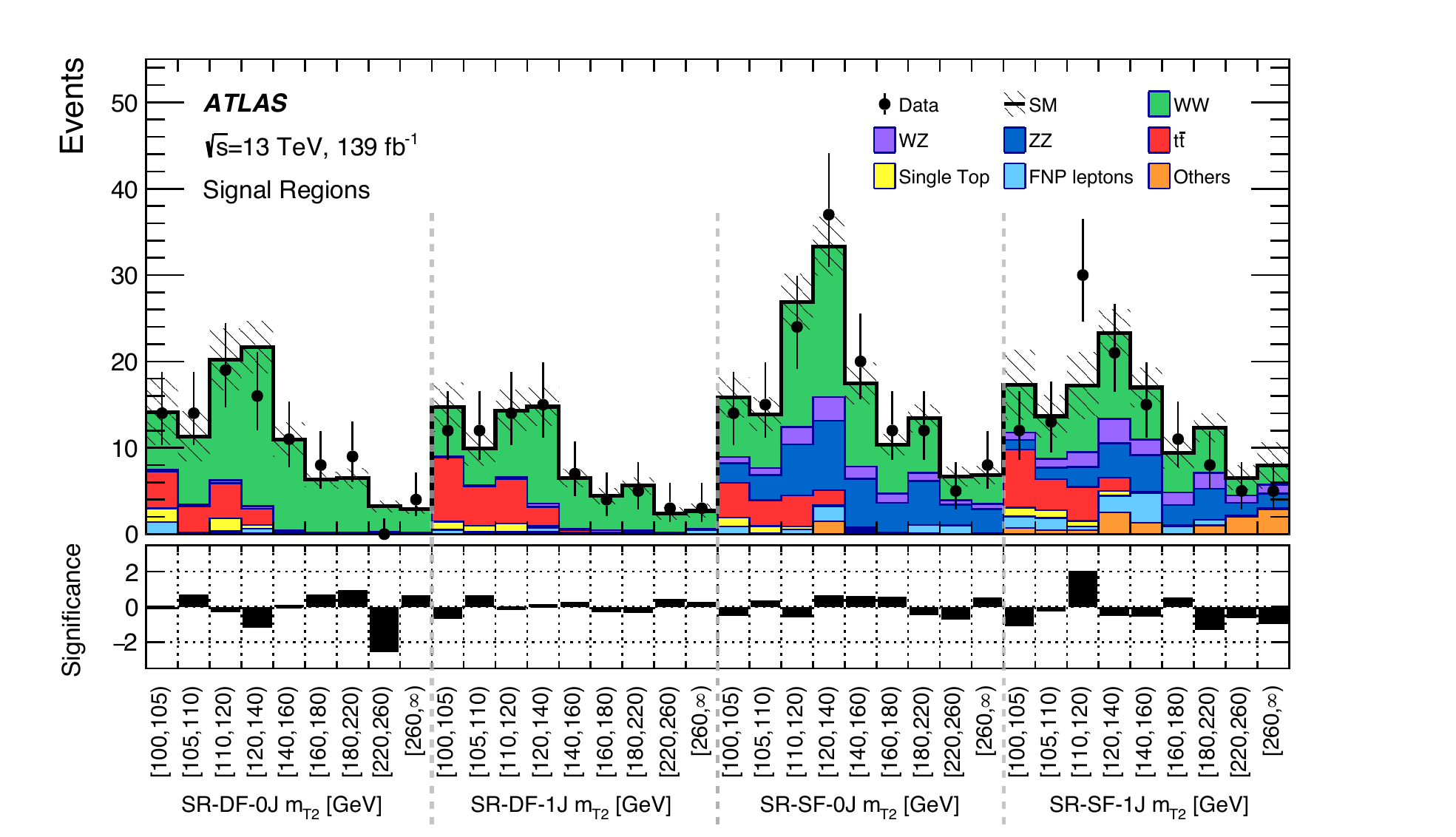}
\caption{The upper panel shows the observed number of events in each of the SRs defined in Table~\ref{tab:SRdefs}, together with the expected SM backgrounds obtained after the background-only fit in the CRs. The shaded band represents the total uncertainty in the expected SM background.
The lower panel shows the significance as defined in Ref.~\cite{Cousin:2008clt}.
}
\label{fig:SRpull}
\end{figure}

\begin{figure}[h]
\centering
\begin{subfigure}[b]{0.48\linewidth}
\includegraphics[width=\linewidth]{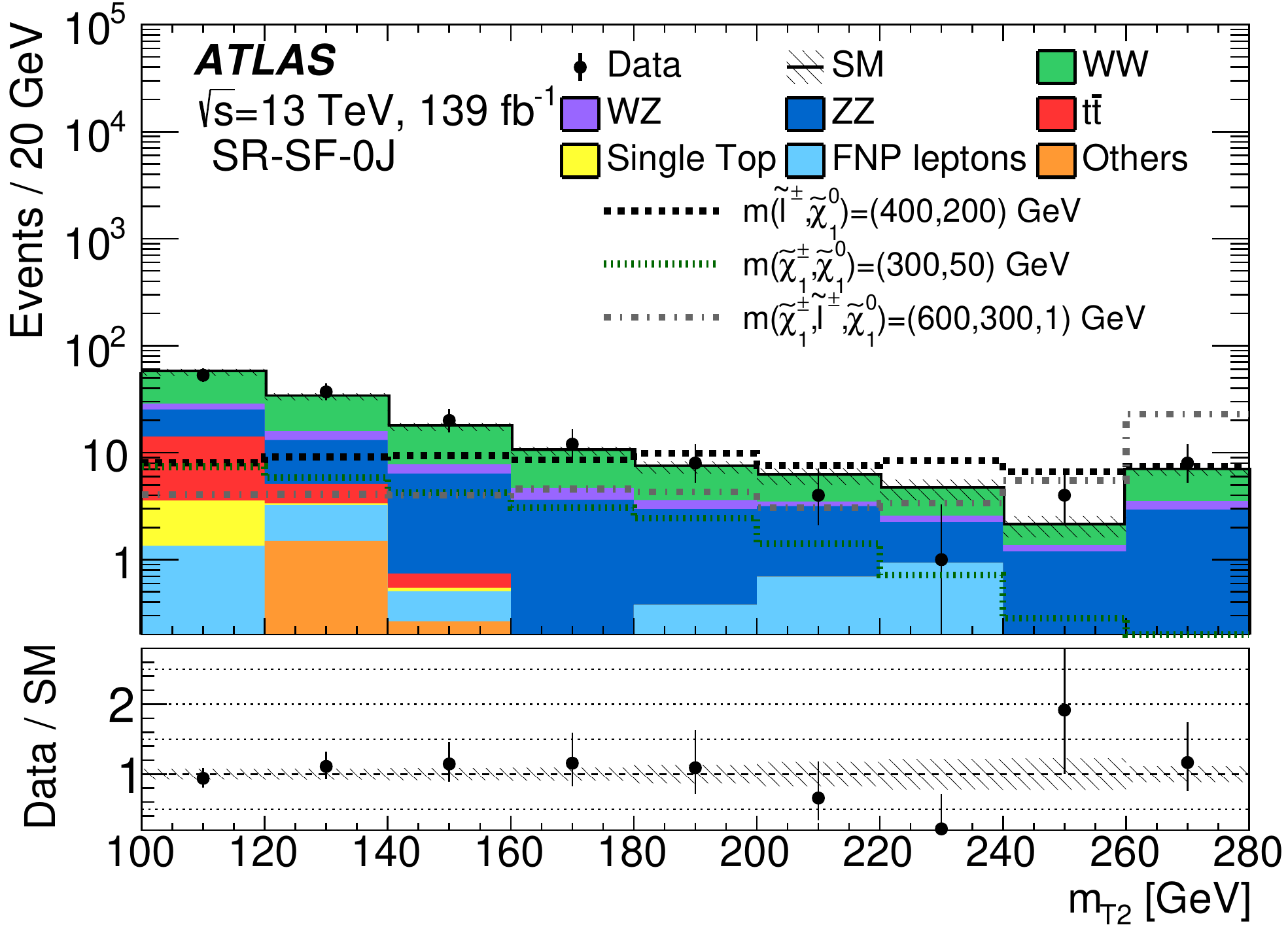}
\caption{\mttwo\ distribution in SR-SF-0J}
\label{figbody:SRSF0_MT2}
\end{subfigure}
\begin{subfigure}[b]{0.48\linewidth}
\includegraphics[width=\linewidth]{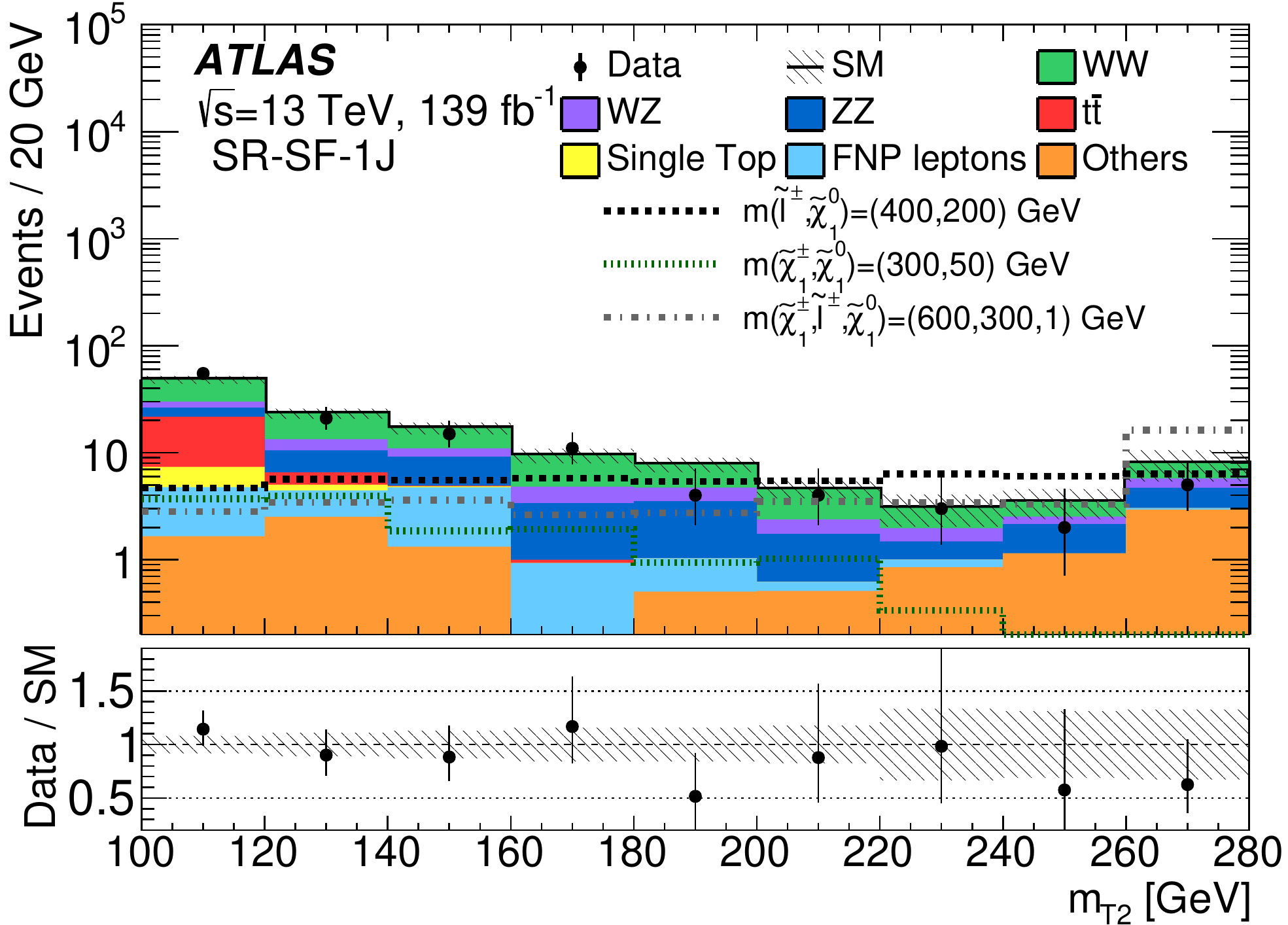}
\caption{\mttwo\ distribution in SR-SF-1J}
\label{figbody:SRSF1_MT2}
\end{subfigure}
\begin{subfigure}[b]{0.48\linewidth}
\includegraphics[width=\textwidth]{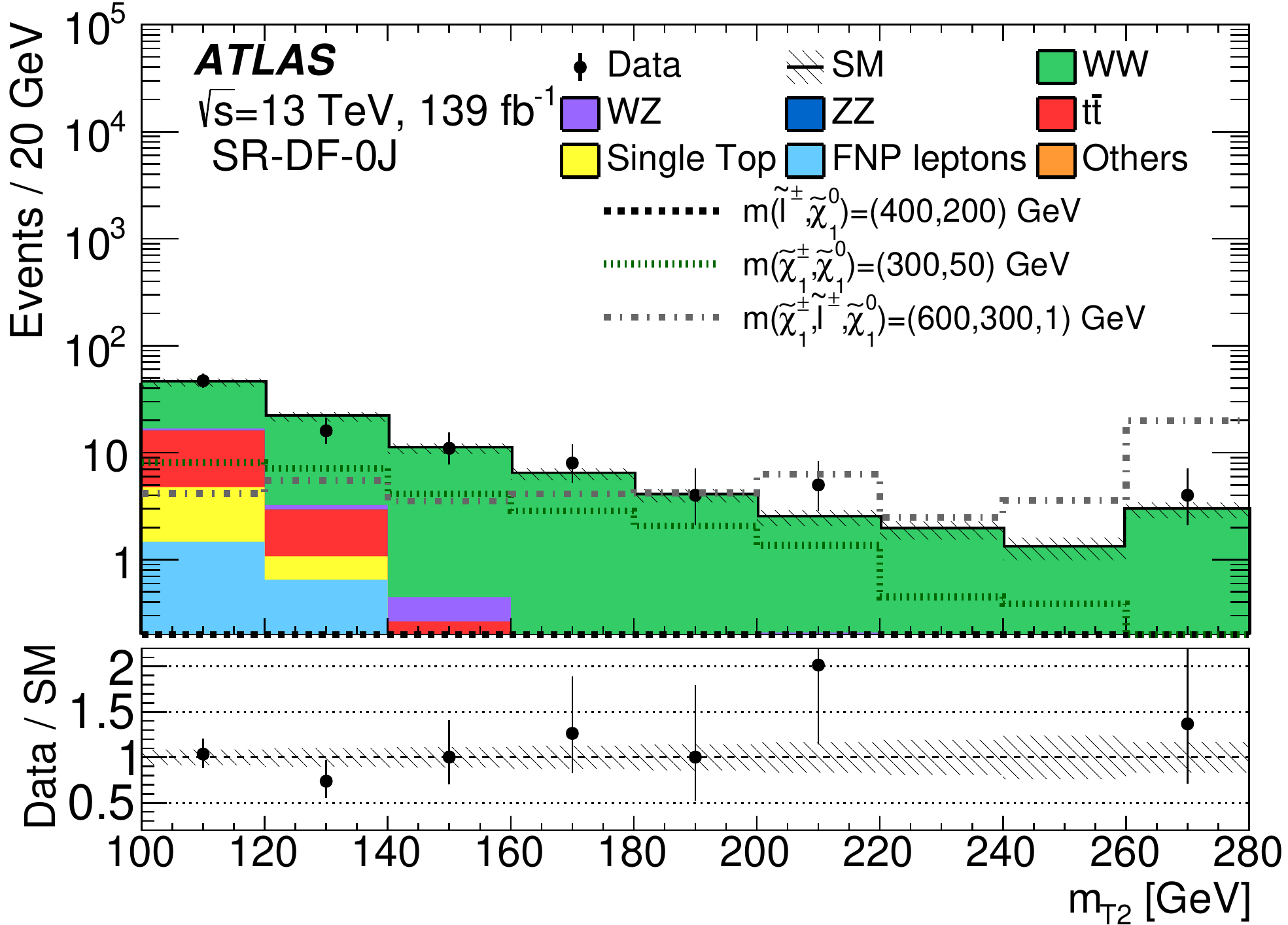}
\caption{\mttwo\ distribution in SR-DF-0J}
\label{figbody:SRDF0_MT2}
\end{subfigure}
\begin{subfigure}[b]{0.48\linewidth}
\includegraphics[width=\linewidth]{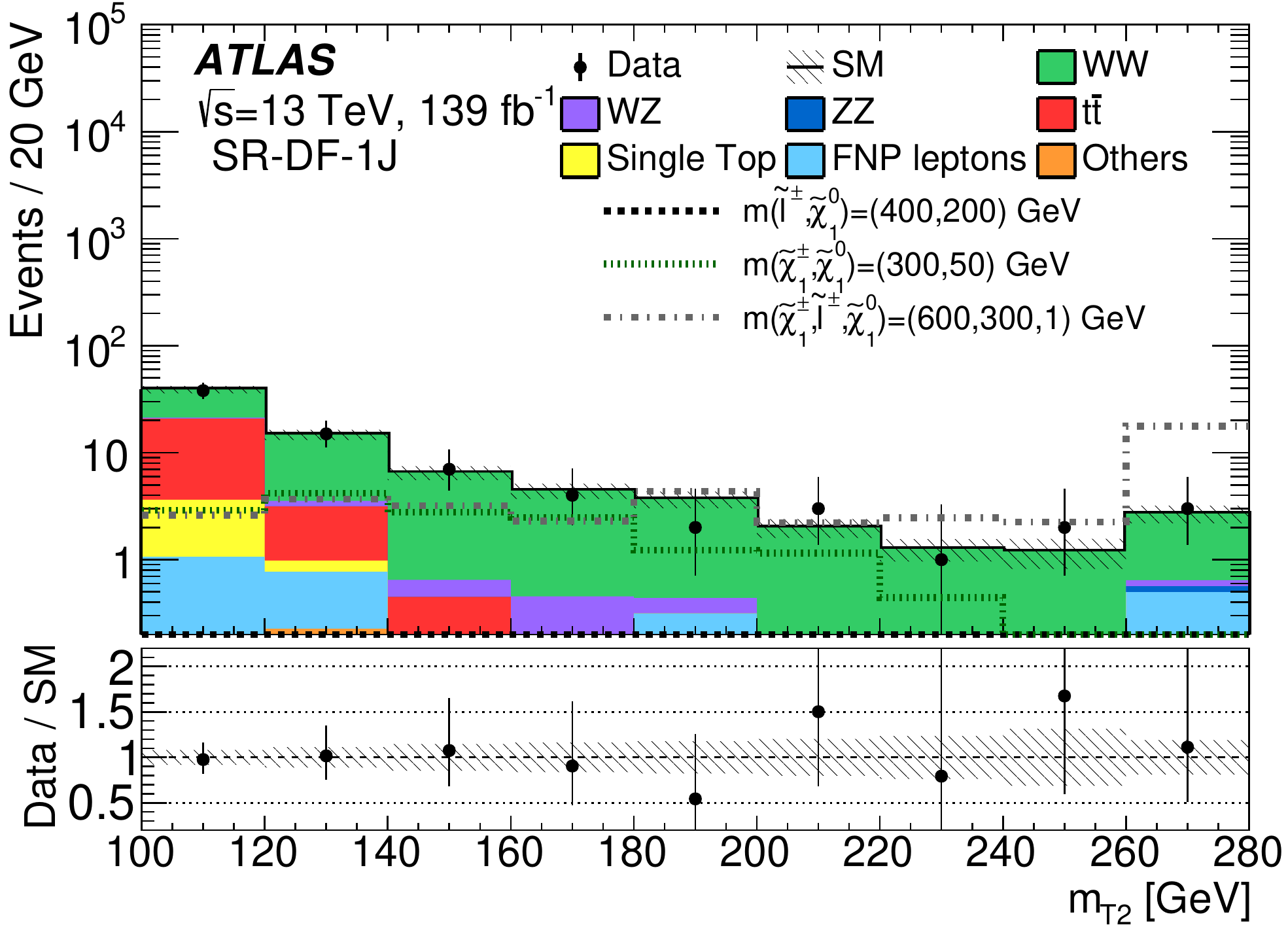}
\caption{\mttwo\ distribution in SR-DF-1J}
\label{figbody:SRDF1_MT2}
\end{subfigure}
\caption{Distributions of \mttwo\ in (a) SR-SF-0J, (b) SR-SF-1J, (c) SR-DF-0J and (d) SR-DF-1J, for data and the estimated SM backgrounds.
The normalisation factors extracted from the corresponding CRs are used to rescale the $t\bar{t}$, single-top-quark, $WW$, $WZ$ and $ZZ$ backgrounds. The FNP lepton background is calculated using the data-driven matrix method. Negligible background contributions are not included in the legends. The uncertainty band includes systematic and statistical errors from all sources and the last bin includes the overflow. Distributions for three benchmark signal points are overlaid for comparison. The lower panels show the ratio of data to the SM background estimate.
}
\label{fig:SRplot}
\end{figure}

The results of the background-only fit in the CRs together with the observed data in the binned SRs are shown in Figure~\ref{fig:SRpull}.
The observed and predicted number of background events in the inclusive SRs are shown in Tables~\ref{tab:SRincDF} and~\ref{tab:SRincSF}. 
Figure~\ref{fig:SRplot} shows the $\mttwo$ distribution for the data and the estimated SM backgrounds for events in the SRs.

\begin{DIFnomarkup}
 
\begin{table}[!htb]
\caption{\label{tab:SRincDF}
Observed event yields and predicted background yields from the fit for the DF inclusive SRs. The model-independent upper limits at 95\% CL on the observed and expected numbers of beyond-the-SM events $S^{0.95}_\mathrm{obs/exp}$ and on the effective beyond-the-SM cross-section 
$\sigma^{0.95}_\mathrm{obs}$ are also shown. The $\pm 1 \sigma $ variations on $S^{0.95}_\mathrm{exp}$ are also provided. The last row shows the $p_0$-value of the SM-only hypothesis. For SRs where the data yield is smaller than expected, the $p_0$-value is capped at 0.50. `Other backgrounds' include the non-dominant background sources, i.e. $t \bar{t}$+$V$, Higgs boson and Drell--Yan events. A `--' symbol indicates that the background contribution is negligible.}
\begin{center}
\begin{tabular}{l|p{0.16\linewidth}p{0.16\linewidth}p{0.16\linewidth}p{0.16\linewidth}}
\hline
Region                 	& SR-DF-0J                  & SR-DF-0J              &  SR-DF-0J                 &  SR-DF-0J             \\
$\mttwo$ [\GeV]	        & $\in$[100,$\infty$)       & $\in$[160,$\infty$)   & $\in$[100,120)            & $\in$[120,160)        \\
\hline
Observed events         & \hspace{5pt} $95$         & $21$                  & \hspace{5pt} $47$         & \hspace{5pt} $27$                  \\
\noalign{\smallskip}\hline\noalign{\smallskip}
Fitted backgrounds      & \hspace{5pt} $96 \pm 15$  & $18.8 \pm 2.4$        & \hspace{5pt} $45 \pm 9$   & \hspace{5pt} $33 \pm 5$            \\
\noalign{\smallskip}\hline\noalign{\smallskip}
{$\!\begin{aligned}
& \text{Fitted } WW         \\[-2pt]
& \text{Fitted } WZ         \\[-2pt]
& \text{Fitted } ZZ         \\[-2pt]
& \text{Fitted } t \bar{t}  \\[-2pt]
& \text{Fitted single top}  \\[-2pt]
& \text{Other backgrounds}  \\[-2pt]
& \text{FNP leptons}
\end{aligned}
$}
&   {$\!\begin{aligned}
76      &\pm 10         \\[-2pt]
1.53    &\pm 0.17       \\[-2pt]
0.00&_{-0.00}^{+0.19}   \\[-2pt]
13      &\pm 7          \\[-2pt]
3.7     &\pm 2.0        \\[-2pt]
0.24    &\pm 0.08       \\[-2pt]
1.8    &\pm 0.6
\end{aligned}
$}
&   {$\!\begin{aligned}
18.2    &\pm 2.4        \\[-2pt]
0.40    &\pm 0.07       \\[-2pt]
0.14    &\pm 0.03       \\[-2pt]
&-              \\[-2pt]
&-              \\[-2pt]
0.07    &\pm 0.02       \\[-2pt]
&-
\end{aligned}
$}
&   {$\!\begin{aligned}
29      &\pm 4          \\[-2pt]
0.66    &\pm 0.11       \\[-2pt]
0.06&_{-0.06}^{+0.23}   \\[-2pt]
11      &\pm 6          \\[-2pt]
3.3     &\pm 1.8        \\[-2pt]
0.08    &\pm 0.02       \\[-2pt]
1.4     &\pm 0.4
\end{aligned}
$}
&   {$\!\begin{aligned}
29      &\pm 4          \\[-2pt]
0.47    &\pm 0.07       \\[-2pt]
<& \text{ }0.04          \\[-2pt]
2.1     &\pm 1.2        \\[-2pt]
0.42    &\pm 0.25       \\[-2pt]
0.09    &\pm 0.05       \\[-2pt]
0.47    &\pm 0.17
\end{aligned}
$} \\
\noalign{\smallskip}\hline\noalign{\smallskip}
{$\!\begin{aligned}
& S^{0.95}_\mathrm{obs}                                     \\[2pt]
& S^{0.95}_\mathrm{exp}                                     \\[2pt]
& {\mathrm \sigma}^{0.95}_\mathrm{obs} \text{ [fb]}         \\[2pt]
& p_0
\end{aligned}
$}
&   {$\!\begin{aligned}
34&.1                   \\[2pt]
35&.2^{+13.9}_{-10.0}   \\[2pt]
0&.24                   \\[2pt]
0&.50
\end{aligned}
$}
&   {$\!\begin{aligned}
12&.7                \\[2pt]
11&.0^{+4.9}_{-3.2}  \\[2pt]
0&.09                \\[2pt]
0&.33
\end{aligned}
$}
&   {$\!\begin{aligned}
23&.8                \\[2pt]
22&.8^{+9.1}_{-6.5}  \\[2pt]
0&.17                \\[2pt]
0&.44
\end{aligned}
$}
&   {$\!\begin{aligned}
11&.8                \\[2pt]
15&.1^{+6.3}_{-4.5}  \\[2pt]
0&.08                \\[2pt]
0&.50
\end{aligned}
$} \\
\noalign{\smallskip}\hline\noalign{\smallskip}
\end{tabular}
 
\vspace{1cm}
 
\begin{tabular}{l|p{0.16\linewidth}p{0.16\linewidth}p{0.16\linewidth}p{0.16\linewidth}}
\hline
Region                 	& SR-DF-1J              & SR-DF-1J                  & SR-DF-1J          & SR-DF-1J              \\
$\mttwo$ [\GeV]	        & $\in$[100,$\infty$)   & $\in$[160,$\infty$)       & $\in$[100,120)    & $\in$[120,160)        \\
\hline
Observed events         & \hspace{5pt} $75$                  & $15$                      & \hspace{5pt} $38$              & $22$                  \\
\noalign{\smallskip}\hline\noalign{\smallskip}
Fitted backgrounds      & \hspace{5pt} $75 \pm 9$            & $15.1 \pm 2.7$            & \hspace{5pt} $39 \pm 6$        & $21.3 \pm 2.8$        \\
\noalign{\smallskip}\hline\noalign{\smallskip}
{$\!\begin{aligned}
& \text{Fitted } WW         \\[-2pt]
& \text{Fitted } WZ         \\[-2pt]
& \text{Fitted } ZZ         \\[-2pt]
& \text{Fitted } t \bar{t}  \\[-2pt]
& \text{Fitted single top}  \\[-2pt]
& \text{Other backgrounds}  \\[-2pt]
& \text{FNP leptons}
\end{aligned}
$}
&   {$\!\begin{aligned}
48      &\pm 8          \\[-2pt]
1.54    &\pm 0.21       \\[-2pt]
0.08    &\pm 0.01       \\[-2pt]
20      &\pm 7          \\[-2pt]
2.8     &\pm 1.4        \\[-2pt]
0.80    &\pm 0.13       \\[-2pt]
2.2     &\pm 0.6
\end{aligned}
$}
&   {$\!\begin{aligned}
13.4    &\pm 2.6        \\[-2pt]
0.53    &\pm 0.12       \\[-2pt]
0.07&_{-0.07}^{+0.24}   \\[-2pt]
0.09    &\pm 0.03       \\[-2pt]
&-              \\[-2pt]
0.25    &\pm 0.05       \\[-2pt]
0.71    &\pm 0.16
\end{aligned}
$}
&   {$\!\begin{aligned}
17.7    &\pm 2.6        \\[-2pt]
0.43    &\pm 0.09       \\[-2pt]
<& \text{ }0.04 \\[-2pt]
17      &\pm 6          \\[-2pt]
2.6     &\pm 1.3        \\[-2pt]
0.19    &\pm 0.10       \\[-2pt]
0.87    &\pm 0.29
\end{aligned}
$}
&   {$\!\begin{aligned}
17.1    &\pm 2.8        \\[-2pt]
0.59    &\pm 0.11       \\[-2pt]
0.01    &\pm 0.00       \\[-2pt]
2.4     &\pm 0.9        \\[-2pt]
0.21    &\pm 0.13       \\[-2pt]
0.34    &\pm 0.04       \\[-2pt]
0.59    &\pm 0.16
\end{aligned}
$} \\
\noalign{\smallskip}\hline\noalign{\smallskip}
{$\!\begin{aligned}
& S^{0.95}_\mathrm{obs}                                     \\[2pt]
& S^{0.95}_\mathrm{exp}                                     \\[2pt]
& {\mathrm \sigma}^{0.95}_\mathrm{obs} \text{ [fb]}         \\[2pt]
& p_0
\end{aligned}
$}
&   {$\!\begin{aligned}
25&.1                   \\[2pt]
25&.3^{+10.3}_{-7.2}   \\[2pt]
0&.18                   \\[2pt]
0&.50
\end{aligned}
$}
&   {$\!\begin{aligned}
10&.2                \\[2pt]
10&.3^{+4.6}_{-3.0}  \\[2pt]
0&.07                \\[2pt]
0&.50
\end{aligned}
$}
&   {$\!\begin{aligned}
16&.8                \\[2pt]
17&.6^{+7.3}_{-5.1}  \\[2pt]
0&.12                \\[2pt]
0&.50
\end{aligned}
$}
&   {$\!\begin{aligned}
12&.3                \\[2pt]
11&.9^{+5.2}_{-3.3}  \\[2pt]
0&.09                \\[2pt]
0&.45
\end{aligned}
$} \\
\noalign{\smallskip}\hline\noalign{\smallskip}
\end{tabular}
\end{center}
\end{table}

\end{DIFnomarkup}

\begin{DIFnomarkup}

\begin{table}[!htb]
\caption{\label{tab:SRincSF}
Observed event yields and predicted background yields from the fit for the SF inclusive SRs. The model-independent upper limits at 95\% CL on the observed and expected numbers of beyond-the-SM events $S^{0.95}_\mathrm{obs/exp}$ and on the effective beyond-the-SM cross-section 
$\sigma^{0.95}_\mathrm{obs}$ are also shown. The $\pm 1 \sigma $ variations on $S^{0.95}_\mathrm{exp}$ are also provided. The last row shows the $p_0$-value of the SM-only hypothesis. For SRs where the data yield is smaller than expected, the $p_0$-value is capped at 0.50. `Other backgrounds' include the non-dominant background sources, i.e. $t \bar{t}$+$V$, Higgs boson and Drell--Yan events. A `--' symbol indicates that the background contribution is negligible.}
\begin{center}
\begin{tabular}{l|p{0.16\linewidth} p{0.16\linewidth} p{0.16\linewidth} p{0.16\linewidth}}
\hline
Region                		& SR-SF-0J      & SR-SF-0J     &  SR-SF-0J   &  SR-SF-0J  \\
$\mttwo$ [\GeV]	& $\in$[100,$\infty$) & $\in$[160,$\infty$) & $\in$[100,120) & $\in$[120,160) \\
\hline
Observed events         & \hspace{0.1pt} $147$          & $37$                      & \hspace{5pt} $53$                  & \hspace{5pt} $57$                    \\
\noalign{\smallskip}\hline\noalign{\smallskip}
Fitted backgrounds      & \hspace{0.1pt} $144 \pm 12$   & $37.3 \pm 3.0$            & \hspace{5pt} $56 \pm 6$            & \hspace{5pt} $51 \pm 5$        \\
\noalign{\smallskip}\hline\noalign{\smallskip}
{$\!\begin{aligned}
& \text{Fitted } WW         \\[-2pt]
& \text{Fitted } WZ         \\[-2pt]
& \text{Fitted } ZZ         \\[-2pt]
& \text{Fitted } t \bar{t}  \\[-2pt]
& \text{Fitted single top}  \\[-2pt]
& \text{Other backgrounds}  \\[-2pt]
& \text{FNP leptons}
\end{aligned}
$}
&   {$\!\begin{aligned}
73      &\pm 8    \\[-2pt]
10.8    &\pm 0.8    \\[-2pt]
38.6    &\pm 2.6   \\[-2pt]
13      &\pm 4    \\[-2pt]
2.4     &\pm 1.4     \\[-2pt]
2.1     &\pm 1.5   \\[-2pt]
5.4     &\pm 1.4
\end{aligned}
$}
&   {$\!\begin{aligned}
18.1    &\pm 2.1        \\[-2pt]
3.08    &\pm 0.27       \\[-2pt]
13.8    &\pm 1.0        \\[-2pt]
&-              \\[-2pt]
&-              \\[-2pt]
0.10&_{-0.10}^{+0.33}   \\[-2pt]
2.2     &\pm 0.4
\end{aligned}
$}
&   {$\!\begin{aligned}
27.6    &\pm 3.0    \\[-2pt]
3.55    &\pm 0.29   \\[-2pt]
11.1    &\pm 0.8    \\[-2pt]
11      &\pm 4      \\[-2pt]
2.2     &\pm 1.3    \\[-2pt]
0.2&_{-0.2}^{+1.4}  \\[-2pt]
1.1     &\pm 0.6
\end{aligned}
$}
&   {$\!\begin{aligned}
27      &\pm 4      \\[-2pt]
4.2     &\pm 0.5    \\[-2pt]
13.7    &\pm 1.5    \\[-2pt]
1.9     &\pm 0.7    \\[-2pt]
0.15    &\pm 0.09   \\[-2pt]
1.76    &\pm 0.30   \\[-2pt]
2.0     &\pm 0.5
\end{aligned}
$} \\
\noalign{\smallskip}\hline\noalign{\smallskip}
{$\!\begin{aligned}
& S^{0.95}_\mathrm{obs}                                     \\[2pt]
& S^{0.95}_\mathrm{exp}                                     \\[2pt]
& {\mathrm \sigma}^{0.95}_\mathrm{obs} \text{ [fb]}         \\[2pt]
& p_0
\end{aligned}
$}
&   {$\!\begin{aligned}
35&.5                \\[2pt]
33&.6^{+13.6}_{-9.3} \\[2pt]
0&.25                \\[2pt]
0&.44
\end{aligned}
$}
&   {$\!\begin{aligned}
14&.3                \\[2pt]
14&.5^{+6.3}_{-4.2}  \\[2pt]
0&.10                \\[2pt]
0&.50
\end{aligned}
$}
&   {$\!\begin{aligned}
17&.8                \\[2pt]
20&.0^{+8.1}_{-5.6}  \\[2pt]
0&.13                \\[2pt]
0&.50
\end{aligned}
$}
&   {$\!\begin{aligned}
23&.5                \\[2pt]
18&.7^{+7.8}_{-5.3}  \\[2pt]
0&.17                \\[2pt]
0&.25
\end{aligned}
$} \\
 
\noalign{\smallskip}\hline\noalign{\smallskip}
\end{tabular}
 
\vspace{1cm}
 
\begin{tabular}{l|p{0.16\linewidth}p{0.16\linewidth}p{0.16\linewidth}p{0.16\linewidth}}
\hline
Region              & SR-SF-1J          & SR-SF-1J                  &  SR-SF-1J             & SR-SF-1J          \\
$\mttwo$ [\GeV]	    & $\in$[100,$\infty$) & $\in$[160,$\infty$)     & $\in$[100,120)        & $\in$[120,160)    \\
\hline
Observed events     & \hspace{0.1pt} $120$             & \hspace{5pt} $29$                  & \hspace{5pt} $55$             & \hspace{5pt} $36$              \\
\noalign{\smallskip}\hline\noalign{\smallskip}
Fitted backgrounds  & \hspace{0.1pt} $124 \pm 12$      & \hspace{5pt} $36 \pm 5$            & \hspace{5pt} $48 \pm 8$       & \hspace{5pt} $40 \pm 4$        \\
\noalign{\smallskip}\hline\noalign{\smallskip}
{$\!\begin{aligned}
& \text{Fitted } WW         \\[-2pt]
& \text{Fitted } WZ         \\[-2pt]
& \text{Fitted } ZZ         \\[-2pt]
& \text{Fitted } t \bar{t}  \\[-2pt]
& \text{Fitted single top}  \\[-2pt]
& \text{Other backgrounds}  \\[-2pt]
& \text{FNP leptons}
\end{aligned}
$}
&   {$\!\begin{aligned}
48      &\pm 6      \\[-2pt]
13.4    &\pm 1.1    \\[-2pt]
22.2    &\pm 1.8    \\[-2pt]
16      &\pm 8      \\[-2pt]
3.3     &\pm 1.7    \\[-2pt]
11.1    &\pm 4.0    \\[-2pt]
10.3    &\pm 1.5
\end{aligned}
$}
&   {$\!\begin{aligned}
14.1    &\pm 2.1        \\[-2pt]
5.2     &\pm 0.6        \\[-2pt]
9.1     &\pm 1.1        \\[-2pt]
0.07&_{-0.07}^{+0.10}   \\[-2pt]
&-              \\[-2pt]
5.6     &\pm 2.1        \\[-2pt]
1.80    &\pm 0.34
\end{aligned}
$}
&   {$\!\begin{aligned}
18.1    &\pm 2.4    \\[-2pt]
3.62    &\pm 0.33   \\[-2pt]
4.8     &\pm 0.5    \\[-2pt]
14      &\pm 7      \\[-2pt]
2.6     &\pm 1.4    \\[-2pt]
1.7&_{-1.7}^{+2.4}  \\[-2pt]
3.1     &\pm 0.6
\end{aligned}
$}
&   {$\!\begin{aligned}
16.0    &\pm 2.2    \\[-2pt]
4.7     &\pm 0.5    \\[-2pt]
8.2     &\pm 0.9    \\[-2pt]
1.6     &\pm 0.8    \\[-2pt]
0.7     &\pm 0.4    \\[-2pt]
3.8     &\pm 1.3    \\[-2pt]
5.3     &\pm 0.7
\end{aligned}
$} \\
\noalign{\smallskip}\hline\noalign{\smallskip}
{$\!\begin{aligned}
& S^{0.95}_\mathrm{obs}                                     \\[2pt]
& S^{0.95}_\mathrm{exp}                                     \\[2pt]
& {\mathrm \sigma}^{0.95}_\mathrm{obs} \text{ [fb]}         \\[2pt]
& p_0
\end{aligned}
$}
&   {$\!\begin{aligned}
30&.6                \\[2pt]
33&.5^{+13.3}_{-9.3} \\[2pt]
0&.22                \\[2pt]
0&.50
\end{aligned}
$}
&   {$\!\begin{aligned}
11&.2                \\[2pt]
15&.3^{+6.5}_{-4.5}  \\[2pt]
0&.08                \\[2pt]
0&.50
\end{aligned}
$}
&   {$\!\begin{aligned}
27&.3                \\[2pt]
21&.9^{+9.0}_{-6.2}  \\[2pt]
0&.19                \\[2pt]
0&.26
\end{aligned}
$}
&   {$\!\begin{aligned}
12&.6                \\[2pt]
15&.5^{+6.5}_{-4.2}  \\[2pt]
0&.09                \\[2pt]
0&.50
\end{aligned}
$} \\
\noalign{\smallskip}\hline\noalign{\smallskip}
\end{tabular}
\end{center}
\end{table}

\end{DIFnomarkup}
 
No significant deviations from the SM expectations are observed in any of the SRs considered, as shown in Figures \ref{fig:SRpull} and \ref{fig:SRplot}.
The CL$_{\text{s}}$ prescription~\cite{Read:2002hq} is used to set model-independent upper limits at 95\% confidence level (CL) on the visible signal cross-section $\sigma^{0.95}_\mathrm{obs}$, defined as the cross-section times acceptance times efficiency, of processes beyond the SM. They are derived in each inclusive SR by performing a fit that includes the observed yield in the SR as a constraint, and a signal yield in the SR as a free parameter of interest.
The observed ($S^{0.95}_\mathrm{obs}$) and expected ($S^{0.95}_\mathrm{exp}$) limits at 95\% CL on the numbers of events from processes beyond the SM in the inclusive SRs defined in Section~\ref{sec:SRs} are calculated. The $p_0$-values, which represent the probability of the SM background alone to fluctuate to the observed number of events or higher, are also provided and are capped at $p_0 = 0.50$. These results are presented in Tables~\ref{tab:SRincDF} and~\ref{tab:SRincSF} for the DF and SF inclusive SRs, respectively.
 
\begin{figure}[htb!]
\centering
\begin{subfigure}[b]{0.48\linewidth}
\includegraphics[width=\linewidth]{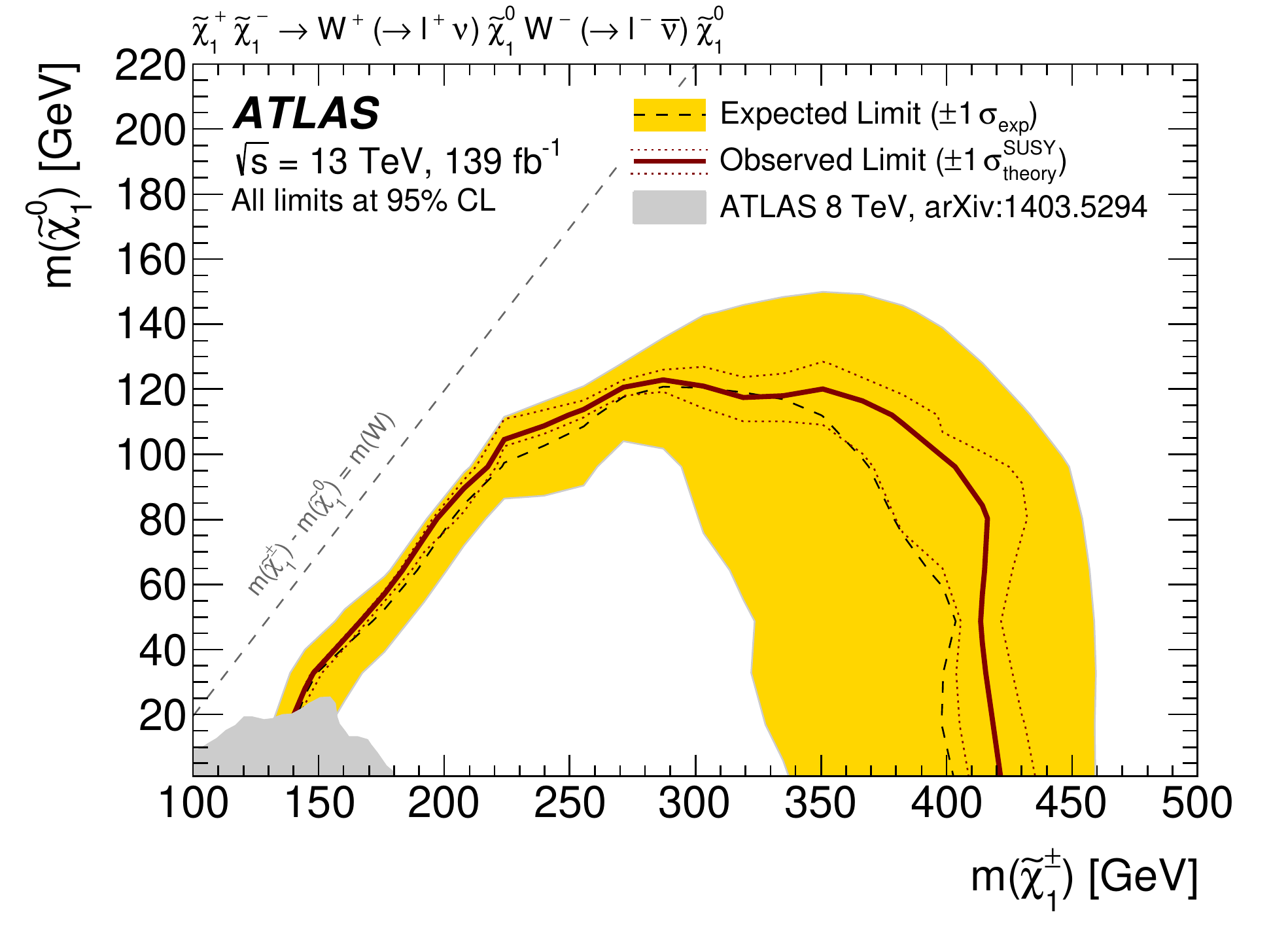}
\caption{$\tilde\chi^+_1\tilde\chi^-_1$ production with $W$-boson-mediated decays}
\label{figbody:C1C1_WW_Limits}
\end{subfigure}
\begin{subfigure}[b]{0.48\linewidth}
\includegraphics[width=\linewidth]{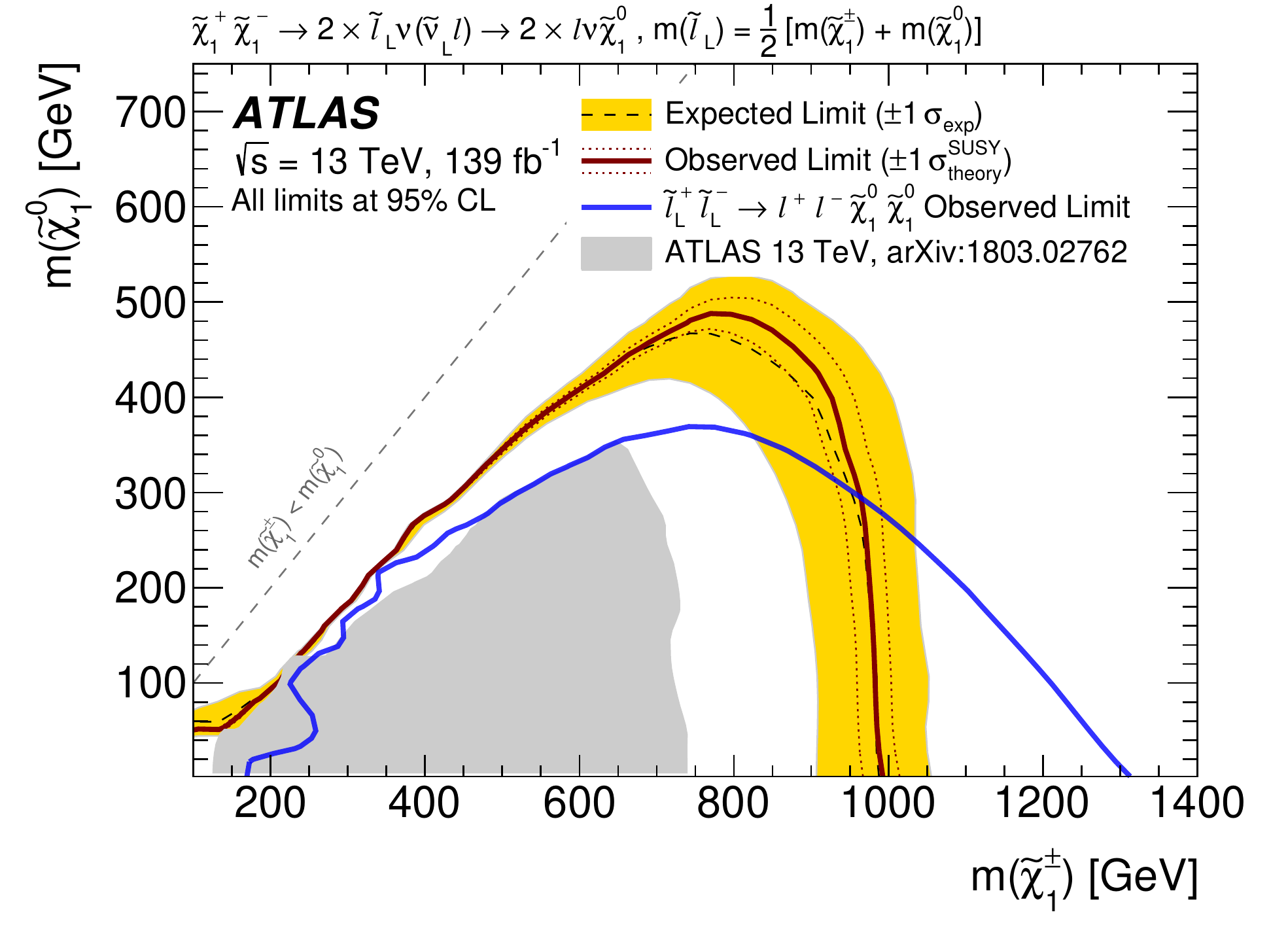}
\caption{$\tilde\chi^+_1\tilde\chi^-_1$ production with $\tilde{\ell}$/$\tilde{\nu}$-mediated decays}
\label{figbody:C1C1_slepslep_Limits}
\end{subfigure}
\begin{subfigure}[b]{0.48\linewidth}
\includegraphics[width=\textwidth]{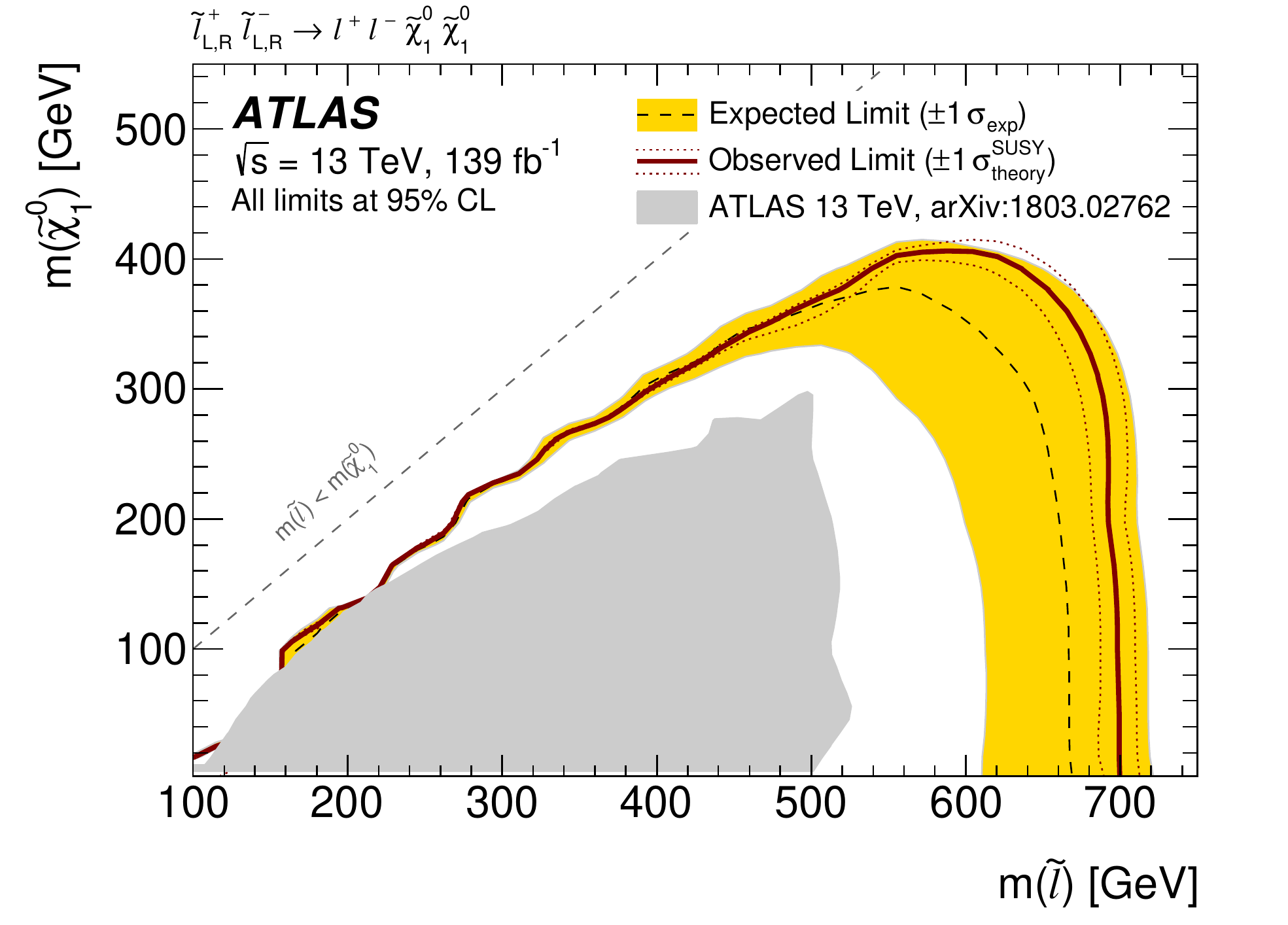}
\caption{$\tilde\ell\tilde\ell$ production}
\label{figbody:SlepSlep_Limits}
\end{subfigure}
 
\caption{
Observed and expected exclusion limits on SUSY simplified models for chargino-pair production with (a) $W$-boson-mediated decays and (b) slepton/sneutrino-mediated decays, and (c) for slepton-pair production. In Figure (b) all three slepton flavours ($\tilde{e}$, $\tilde{\mu}$, $\tilde{\tau}$) are considered, while only $\tilde{e}$ and $\tilde{\mu}$ are considered in Figure (c). The observed (solid thick line) and expected (thin dashed line) exclusion contours are indicated. The upper shaded band corresponds to the $\pm 1 \sigma $ variations in the expected limit, including all uncertainties except theoretical uncertainties in the signal cross-section. The dotted lines around the observed limit illustrate the change in the observed limit as the nominal signal cross-section is scaled up and down by the theoretical uncertainty. The blue line in (b) corresponds to the observed limit for $\tilde{\ell}_{\textup{L}}$ projected into this model for the chosen slepton mass hypothesis (slepton masses midway between the mass of the chargino and that of the $\tilde{\chi}_1^0$). All limits are computed at 95\% CL\@. The observed limits obtained by ATLAS in previous 
searches are also shown (lower shaded areas)~\cite{SUSY-2013-11,SUSY-2016-24}.
}
\label{fig:exclusion}
\end{figure}

\begin{figure}[htb!]
\centering
\begin{subfigure}[b]{0.48\linewidth}
\includegraphics[width=\linewidth]{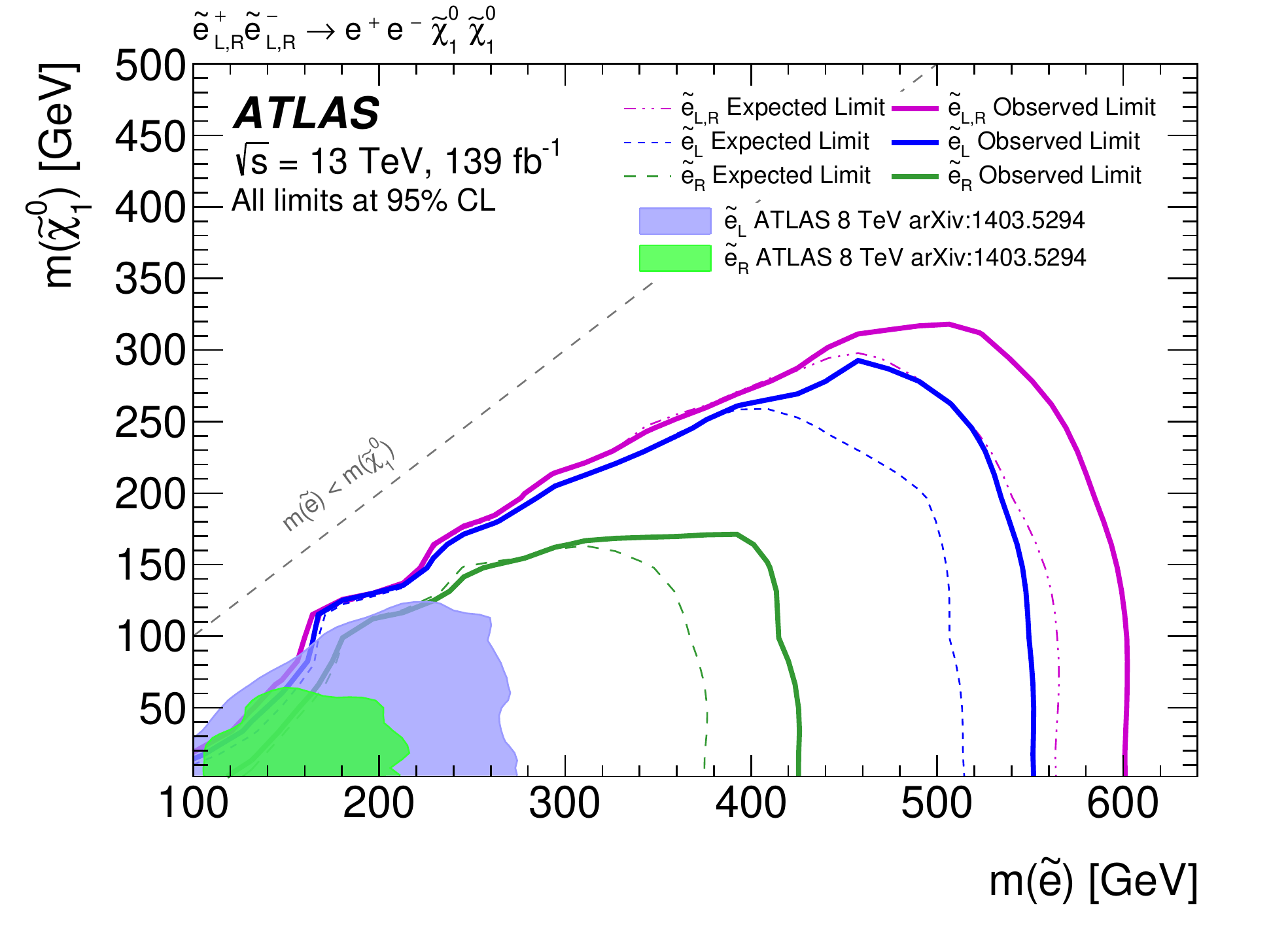}
\caption{$\tilde{e}\tilde{e}$ production}
\label{figbody:C1C1_WW_Limits}
\end{subfigure}
\begin{subfigure}[b]{0.48\linewidth}
\includegraphics[width=\linewidth]{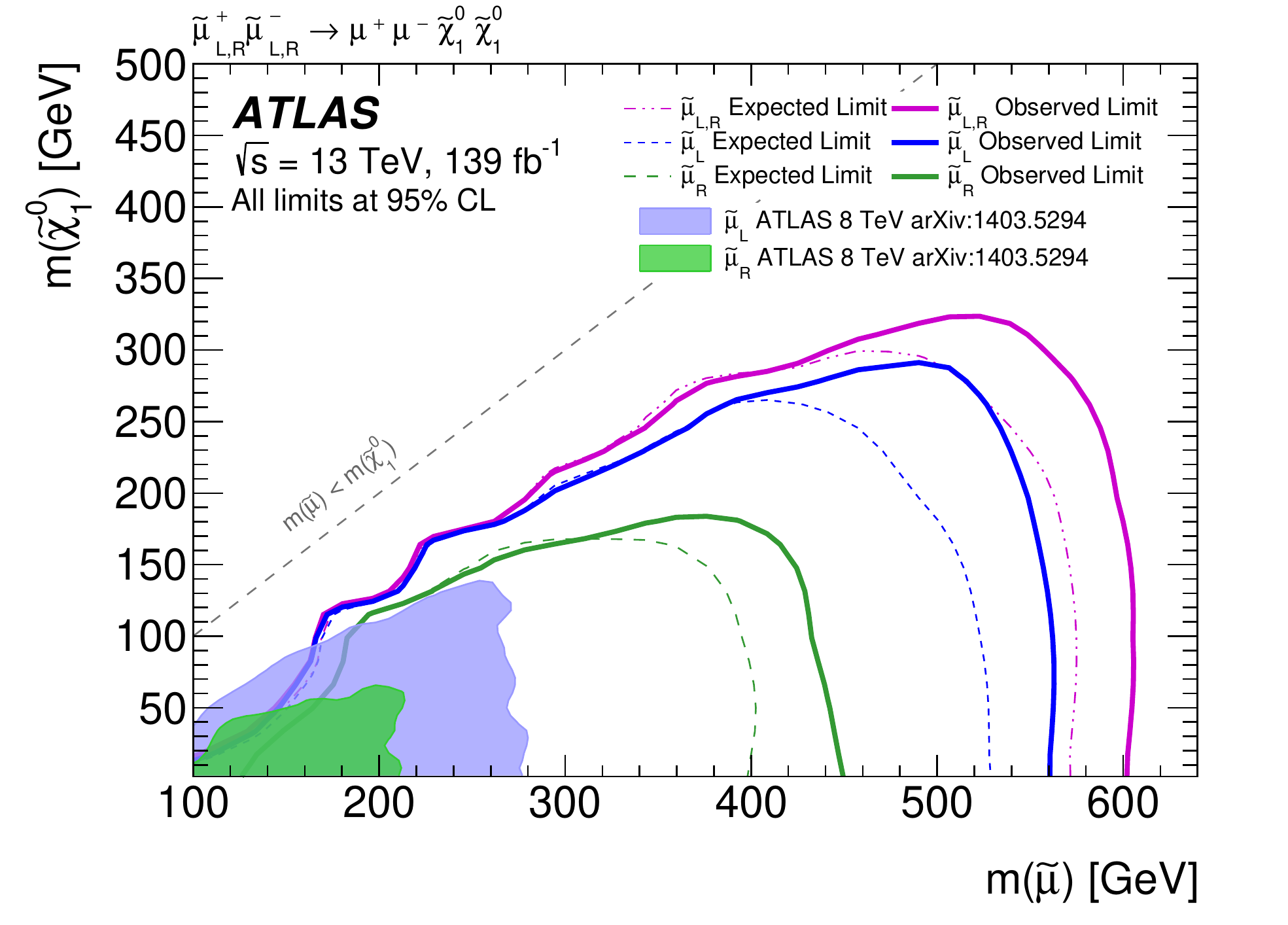}
\caption{$\tilde{\mu}\tilde{\mu}$}
\label{figbody:C1C1_slepslep_Limits}
\end{subfigure}
 
\caption{
Observed and expected exclusion limits on SUSY simplified models for (a) direct selectron production and (b) direct smuon production.
In Figure (a) the observed (solid thick lines) and expected (dashed lines) exclusion contours are indicated for combined $\tilde{e}_{\textup{L,R}}$ and for $\tilde{e}_{\textup{L}}$ and $\tilde{e}_{\textup{R}}$.
In Figure (b) the observed (solid thick lines) and expected (dashed lines) exclusion contours are indicated for combined $\tilde{\mu}_{\textup{L,R}}$ and for $\tilde{\mu}_{\textup{L}}$ and $\tilde{\mu}_{\textup{R}}$. All limits are computed at 95\% CL.\@
The observed limits obtained by ATLAS in previous 
searches are also shown in the shaded areas~\cite{SUSY-2016-24}.
}
\label{fig:sleptonExclusions}
\end{figure}

Exclusion limits at 95\% CL are set on the masses of the chargino, neutralino and sleptons for the simplified models  shown in Figure \ref{fig:c1c1viawwfeynman}. These also use the CL$_{\text{s}}$ prescription and include the exclusive SRs and the CRs in the simultaneous likelihood fit.  For the models of chargino pair production the SF and DF SRs are included in the likelihood fit, whilst for direct slepton production only the SF SRs are included. The results are shown in Figure~\ref{fig:exclusion}. In the model of direct chargino pair production with decays via $W$ bosons with a massless $\ninoone$, $\chinoonepm$ masses up to 420~\GeV\ are excluded at 95\% CL\@.
In the model of direct chargino pair production with decays via sleptons or sneutrinos with a massless $\ninoone$, $\chinoonepm$ masses up to 1~\TeV\ are excluded at 95\% CL\@. Finally, in the model of direct slepton pair production with a massless $\ninoone$, slepton masses up to 700~\GeV\ are excluded at 95\% CL\@. For direct slepton production, exclusion limits are also set for selectrons and smuons separately by including only the di-electron and di-muon SF SRs in the likelihood fit respectively. These are shown in Figure \ref{fig:sleptonExclusions} for single slepton species
$\tilde{e}_{\textup{R}}$, $\tilde{\mu}_{\textup{R}}$, $\tilde{e}_{\textup{L}}$, $\tilde{\mu}_{\textup{L}}$ along with combined limits for mass-degenerate $\tilde{e}_{\textup{L,R}}$ and $\tilde{\mu}_{\textup{L,R}}$. The observed limit for $\tilde{\ell}_{\textup{L}}$ is shown on the exclusion plot for chargino pair production with slepton-mediated decays in Figure \ref{fig:exclusion} for comparison. However since the sensitivity does not depend strongly on the slepton mass hypothesis for a broad range of slepton masses~\cite{SUSY-2013-11}, these results are applicable for many models not excluded by the direct slepton limits. These results significantly extend the previous exclusion limits~\cite{SUSY-2013-11,SUSY-2016-24,CMS-SUS-13-006,CMS-SUS-14-002,CMS-SUS-16-039,CMS-SUS-17-004,CMS-SUS-17-010} for the same scenarios.

\FloatBarrier

\section{Conclusion}
\label{sec:conclusions}
 
A search for the electroweak production of charginos and sleptons decaying into final states with exactly two oppositely charged leptons and missing transverse momentum is presented. The analysis uses 139~fb$^{-1}$ of $\sqrt{s}=13$~\TeV\ proton--proton collisions recorded by the ATLAS detector at the LHC between 2015 and 2018.
Three scenarios are considered: the production of lightest-chargino pairs, followed by their decays into final states with leptons and the lightest neutralino via either $W$ bosons or sleptons/sneutrinos, and direct production of slepton pairs, where each slepton decays directly into the lightest neutralino and a lepton and
different assumptions about the masses of the superpartners of the left-handed and right-handed charged leptons, $\tilde{e}_\mathrm{L}$, $\tilde{e}_\mathrm{R}$, $\tilde{\mu}_\mathrm{L}$ and $\tilde{\mu}_\mathrm{R}$, 
are considered.
No significant deviations from the Standard Model expectations are observed and limits at 95\% CL are set on the masses of relevant supersymmetric particles in each of these scenarios. For a massless lightest neutralino, masses up to 420~\GeV\ are excluded for the production of the lightest-chargino pairs assuming $W$-boson-mediated decays and up to 1~\TeV\ for slepton-pair-mediated decays, whereas for slepton-pair production masses up to 700~\GeV\ are excluded assuming three generations of mass-degenerate sleptons.
These results significantly extend the previous exclusion limits for the same scenarios.

\section*{Acknowledgements}
 
% The next lines are included from the .//acknowledgements/Acknowledgements.tex input file
 
We thank CERN for the very successful operation of the LHC, as well as the
support staff from our institutions without whom ATLAS could not be
operated efficiently.
 
We acknowledge the support of ANPCyT, Argentina; YerPhI, Armenia; ARC, Australia; BMWFW and FWF, Austria; ANAS, Azerbaijan; SSTC, Belarus; CNPq and FAPESP, Brazil; NSERC, NRC and CFI, Canada; CERN; CONICYT, Chile; CAS, MOST and NSFC, China; COLCIENCIAS, Colombia; MSMT CR, MPO CR and VSC CR, Czech Republic; DNRF and DNSRC, Denmark; IN2P3-CNRS, CEA-DRF/IRFU, France; SRNSFG, Georgia; BMBF, HGF, and MPG, Germany; GSRT, Greece; RGC, Hong Kong SAR, China; ISF and Benoziyo Center, Israel; INFN, Italy; MEXT and JSPS, Japan; CNRST, Morocco; NWO, Netherlands; RCN, Norway; MNiSW and NCN, Poland; FCT, Portugal; MNE/IFA, Romania; MES of Russia and NRC KI, Russian Federation; JINR; MESTD, Serbia; MSSR, Slovakia; ARRS and MIZ\v{S}, Slovenia; DST/NRF, South Africa; MINECO, Spain; SRC and Wallenberg Foundation, Sweden; SERI, SNSF and Cantons of Bern and Geneva, Switzerland; MOST, Taiwan; TAEK, Turkey; STFC, United Kingdom; DOE and NSF, United States of America. In addition, individual groups and members have received support from BCKDF, CANARIE, CRC and Compute Canada, Canada; COST, ERC, ERDF, Horizon 2020, and Marie Sk{\l}odowska-Curie Actions, European Union; Investissements d' Avenir Labex and Idex, ANR, France; DFG and AvH Foundation, Germany; Herakleitos, Thales and Aristeia programmes co-financed by EU-ESF and the Greek NSRF, Greece; BSF-NSF and GIF, Israel; CERCA Programme Generalitat de Catalunya, Spain; The Royal Society and Leverhulme Trust, United Kingdom.
 
The crucial computing support from all WLCG partners is acknowledged gratefully, in particular from CERN, the ATLAS Tier-1 facilities at TRIUMF (Canada), NDGF (Denmark, Norway, Sweden), CC-IN2P3 (France), KIT/GridKA (Germany), INFN-CNAF (Italy), NL-T1 (Netherlands), PIC (Spain), ASGC (Taiwan), RAL (UK) and BNL (USA), the Tier-2 facilities worldwide and large non-WLCG resource providers. Major contributors of computing resources are listed in Ref.~\cite{ATL-GEN-PUB-2016-002}.
 
% End of text imported from the .//acknowledgements/Acknowledgements.tex input file

\clearpage
\printbibliography
\begin{DIFnomarkup}

\end{DIFnomarkup}
\clearpage % ATLAS Collaboration author list
% Reference date of SUSY-2018-32 is 2019-03-05
% Author list last updated on date 22-FEB-20
% Data extracted on 22-Feb-2020 for paper reference SUSY-2018-32
% at 1:16pm
 
\begin{flushleft}
{\Large The ATLAS Collaboration}

\bigskip

G.~Aad$^\textrm{\scriptsize 102}$,    
B.~Abbott$^\textrm{\scriptsize 129}$,    
D.C.~Abbott$^\textrm{\scriptsize 103}$,    
O.~Abdinov$^\textrm{\scriptsize 13,*}$,    
A.~Abed~Abud$^\textrm{\scriptsize 71a,71b}$,    
K.~Abeling$^\textrm{\scriptsize 53}$,    
D.K.~Abhayasinghe$^\textrm{\scriptsize 94}$,    
S.H.~Abidi$^\textrm{\scriptsize 167}$,    
O.S.~AbouZeid$^\textrm{\scriptsize 40}$,    
N.L.~Abraham$^\textrm{\scriptsize 156}$
% ...
\bigskip
\\

% ...
$^{13}$Institute of Physics, Azerbaijan Academy of Sciences, Baku; Azerbaijan.\\
% ...
$^{40}$Niels Bohr Institute, University of Copenhagen, Copenhagen; Denmark.\\
% ...
$^{53}$II. Physikalisches Institut, Georg-August-Universit\"{a}t G\"ottingen, G\"ottingen; Germany.\\
% ...
$^{71}$$^{(a)}$INFN Sezione di Pavia;$^{(b)}$Dipartimento di Fisica, Universit\`a di Pavia, Pavia; Italy.\\
% ...
$^{94}$Department of Physics, Royal Holloway University of London, Egham; United Kingdom.\\
% ...
$^{102}$CPPM, Aix-Marseille Universit\'e, CNRS/IN2P3, Marseille; France.\\
$^{103}$Department of Physics, University of Massachusetts, Amherst MA; United States of America.\\
% ...
$^{129}$Homer L. Dodge Department of Physics and Astronomy, University of Oklahoma, Norman OK; United States of America.\\
% ...
$^{156}$Department of Physics and Astronomy, University of Sussex, Brighton; United Kingdom.\\
% ...
$^{167}$Department of Physics, University of Toronto, Toronto ON; Canada.\\
% ...
$^{a}$ Also at Borough of Manhattan Community College, City University of New York, New York NY; United States of America.\\
$^{b}$ Also at CERN, Geneva; Switzerland.\\
% ...
$^{*}$ Deceased

\end{flushleft}

% Created with Glance <Atlas.Glance@cern.ch>

\end{document}